\documentclass[a4paper,11pt]{article}
\usepackage{jcappub} % for details on the use of the package, please
\usepackage[T1]{fontenc} % if needed
\usepackage[dvipsnames]{xcolor}
\usepackage{slashed}
\usepackage{subcaption}
\usepackage{physics}
\usepackage{csquotes}
\usepackage{caption}

\usepackage{amsmath}
\usepackage{amssymb}
\usepackage{bbm}
\usepackage{hyperref}
\usepackage{dsfont}
\newcommand{\ignore}[1]{}
\usepackage{amsmath,bm}
\usepackage{framed}

\title{\huge \boldmath Modular Invariant Slow Roll Inflation}

\author[a]{Gui-Jun Ding}
\author[a]{Si-Yi Jiang}
\author[b]{Wenbin Zhao}

\affiliation[a]{Department of Modern Physics, University of Science and Technology of China\\
Hefei, Anhui 230026, China}
\affiliation[b]{Bethe Center for Theoretical Physics and Physikalisches Institut,
  Universität Bonn\\
Nussallee 12, 53115 Bonn, Germany}

\emailAdd{dinggj@ustc.edu.cn}
\emailAdd{siichiang@mail.ustc.edu.cn}
\emailAdd{wenbin.zhao@uni-bonn.de}

\abstract{We propose new classes of inflation models based on the modular symmetry, where the modulus field $\tau$ serves as the inflaton. We establish a connection between modular inflation and modular stabilization, wherein the modulus field rolls towards a fixed point along the boundary of the fundamental domain. We find the modular symmetry strongly constrain the possible shape of the potential and identify some parameter space where the inflation predictions agree with cosmic microwave background observations. The tensor-to-scalar ratio is predicted to be smaller than $10^{-6}$ in our models, while the running of spectral index is of the order of $10^{-4}$.}

\begin{document}
\maketitle
\flushbottom

\section{Introduction}

The precise measurements of the cosmic microwave background (CMB)~\cite{Planck:2018vyg,BICEP:2021xfz} strongly favor an exponentially expanding period of early universe called inflation~\cite{Guth:1980zm}. The quantum fluctuations at the beginning of inflation, stretched to cosmologically large scales, became the seeds of the universe's stars and galaxies. The detailed particle physics mechanism responsible for inflation is still unknown, and the hypothetical field thought to be responsible for inflation is called the inflaton. Various inflation theories have been proposed that make radically different predictions, we refer the readers to the comprehensive cosmological monographs~\cite{weinberg2008cosmology,Baumann:2022mni}. The cosmic inflation can be naturally realized, if the inflaton is a scalar field and it slowly rolls over a flat plateau of its potential. This is the well-known slow roll (SR) inflation mechanism~\cite{Linde:1981mu,Albrecht:1982wi}. The characteristic features of CMB, such as the spectral index $n_s$ and tensor-to-scalar ratio $r$, are intricately linked to the shape of the inflation potential.

Modular symmetry and moduli fields (modulus) are ubiquitous in higher dimensional theory such as superstring theory~\cite{Polchinski:1998rq,Becker:2006dvp,Ibanez:2012zz}. The modulus generally denote the scalar degrees of freedom in the effective action of the four dimensional spacetime and describe low energy excitations in the extra dimensions such as the shape and size of the extra compact space. Moreover, the modular symmetry is of great importance is in two-dimensional conformal field theory~\cite{ketov1995conformal,Rychkov:2016iqz,Gillioz:2022yze}. More applications of modular symmetry in physics and mathematics can be found in~\cite{DHoker:2022dxx}. In the torus or orbifold compactification, the shape of the torus is determined by the complex modulus $\tau$. The Yukawa couplings arise from the overlap integral of the zero mode profiles of the matter fields, and they are functions of $\tau$ transforming nontrivially under the modular symmetry group. The modulus $\tau$ and modular flavor symmetry are used to explain the flavor structure of quarks and leptons~\cite{Feruglio:2017spp}, see Refs.~\cite{Kobayashi:2023zzc,Ding:2023htn} for reviews.

The modulus and modular symmetry could have phenomenological implications in cosmology, and the modulus can be a candidate for the inflaton to realize inflation in the early Universe~\cite{Cicoli:2023opf}. It has been showed that successful inflation can not be realized  with a single moduli field for the logarithmic K\"ahler potential~\cite{Badziak:2008yg,Ben-Dayan:2008fhy,Covi:2008cn}. A stabilizer field $X$ besides the modulus $\tau$ was introduced in Refs.~\cite{Kobayashi:2016mzg,Abe:2023ylh} to build an inflation potential, and the K\"ahler potential modified by $X$ was considered to flatten the scalar potential in the whole complex plane. Higher powers of $\tau$ are included in the logarithm of the K\"ahler potential to realize modular inflation in~\cite{Frolovsky:2024xet_pb}. The modular invariance puts strong constraints on the scalar potential of the modulus~\cite{Schimmrigk:2014ica,Schimmrigk:2015qju,Schimmrigk:2016bde}, and the Starobinsky like model can arise in the context of modular symmetry~\cite{Casas:2024jbw}.  

The modular symmetry is broken by the vacuum expectation value (VEV) of the modulus field $\tau$, and there is no VEV of $\tau$ which preserves the full modular symmetry group. Only after the complex modulus $\tau$ obtains a VEV, the modular forms and the Yukawa couplings get fixed. Hence the VEV of the complex modulus $\tau$  has to be dynamically stabilized. The extrema of the modular invariant scalar potentials tend to be close to the boundary of fundamental domain or  the imaginary axis~\cite{Cvetic:1991qm,Kobayashi:2019xvz,Kobayashi:2019uyt,Ishiguro:2020tmo,Novichkov:2022wvg,Leedom:2022zdm,Knapp-Perez:2023nty,King:2023snq,Kobayashi:2023spx,Higaki:2024jdk}. Particularly the fixed points $\tau=i,\omega$ are preferred extrema. Recently it is shown that either Minkowski minima or De Sitter (dS) minima at the fixed points can be achieved by considering non-perturbative corrections to the dilaton K\"ahler potential~\cite{Cvetic:1991qm,Leedom:2022zdm,King:2023snq}. The scalar potential of the modulus can not only dynamically fix its VEV, but can also possibly accommodate inflation. It has been noticed that realizing slow roll inflation in the moduli sector is closely related to admitting metastable dS vacua~\cite{Covi:2008cn}. 

The purpose of this paper is to investigate whether the supergravity motivated potential~\cite{Cvetic:1991qm} for modular stabilization, including  dilation and non-perturbative effects, could also realize slow roll inflation driven by the modulus $\tau$. Motivated by the vacuum structure of $\tau$ and the  special properties of the fixed points $i,\omega=e^{i 2\pi/3},i\infty$ under modular transformation, we shall consider inflationary trajectories along the boundary of fundamental domain, moving from one fixed point to another. After inflation, the modulus stays around the fixed point, which can be used to address the flavor mixing and fermion mass hierarchy in the SM~\cite{Feruglio:2021dte,Novichkov:2021evw,Petcov:2022fjf,Kikuchi:2023cap,Abe:2023ilq,Kikuchi:2023jap,Abe:2023qmr,Petcov:2023vws,Abe:2023dvr,deMedeirosVarzielas:2023crv,Ding:2024xhz}. We find that the scalar potential is strongly constrained by modular symmetry and a class of inflation potentials can be generated in this approach. The resulting predictions are all compatible with current cosmological observations. 
It is promising that the modular symmetry approach can help to explain the flavour puzzle and cosmic inflation in a coherent way.

The remainder of this paper is organised as follows. In section~\ref{sec:framework},  we introduce the setup of our model,  the explicit form of the scalar potential is presented, and we emphasize its special features due to modular invariance. We study the slow roll inflation of the modulus $\tau$ rolling along the boundary of the fundamental domain in section~\ref{sec:MII}. For the case that inflationary trajectory lies on the unit arc in the $\tau$ plane, the quadratic polynomial function $\mathcal{P}(j(\tau))=1+\beta\left(1-\frac{j(\tau)}{1728}\right)+\gamma\left(1-\frac{j(\tau)}{1728}\right)^2$ is considered. Either of $\beta$ or $\gamma$ should be non-zero in order to reproduce the observed value of the spectral index. The ultra slow roll inflation could be realized if the inflaton $\tau$ rolls along the vertical boundary of fundamental domain. Section~\ref{sec:SaC} contains a summary of our result and discussions of some possible further developments. We present the basic aspects of the modular symmetry as well as some relevant formulae of modular forms in Appendix~\ref{app:modularforms}. The slow roll inflation parameters in terms of the scalar potential are collected in Appendix~\ref{app:SRconfig}. We present the numerical results for some examples of modular inflation models in Appendix~\ref{app:MR}. Finally, we briefly discuss the stabilization of dilaton in Appendix~\ref{app:dilatonstablization}.

%%%%%%%%%%%%%%%%%%%%%%%%%%%%%%%%%%%%%%%%%%%%%%%%%%%%%%%%%%%%%%%%%%%%%%%%%%
\section{The framework\label{sec:framework}}

Supergravity is the low-energy limit of superstring theory which is a promising candidate of quantum gravity, and it is  predictive framework allowing to address both inflation and beyond standard model physics. Consequently we adopt the framework of supergravity in this work. In order to be more general, we do not use any concrete models of string theory and any specific compactification mechanism. In the following, we  present the superpotential and K\"ahler potential from which the modular invariant scalar potential is derived. Then the properties of the scalar potential and the constraints on the parameter space of the scalar potential are explored.

%%%%%%%%%%%%%%%%%%%%%%%%%%%%%%%%%%%%%%%%%%%%%%%%%%%%%%%%%%%%%%%%%%%%%%%%%%

\subsection{Modular invariant scalar potential in supergravity\label{subsec:scalar-potential}}

As an effective theory of superstring theory, the spectrum of a $N=1$ supergravitry theory normally contains the dilaton, K\"ahler moduli, complex structure moduli, gauge fields, and twisted and untwisted matter fields after heterotic orbifold compactifications. Given the fact that a single K\"ahler moduli is not enough to realise inflation~\cite{Covi:2008cn,Ben-Dayan:2008fhy}, we find it is natural to include dilaton field into our analysis.  This choice has been adopted in existing literature to study different phenomenon ~\cite{Cvetic:1991qm,Leedom:2022zdm}. In this paper, the K\"ahler modulus field $\tau$ plays the role of inflaton field. The effective action is determined by the following modular-invariant full SUGRA K\"ahler potential~\cite{Ibanez:2012zz,Cerdeno:1998hs}
\begin{equation}\label{MIG}
{\cal G}(\tau,\bar{\tau},S,\bar{S})=\kappa^2{\cal K}(\tau,\bar{\tau},S,\bar{S}) + \ln |\kappa^3{\cal W}(\tau,S)|^2\,,
\end{equation}
where $\kappa=\sqrt{8\pi G_N}=1/M_{\text{Pl}}$ is the
gravitational coupling constant and $M_{\text{Pl}}=2.4\times 10^{18}$ GeV denotes the reduced Planck scale. The field $S$ is the dilaton and it is a modular invariant field. ${\cal G}$ is a combination of the  K\"ahler potential ${\cal K}$ and the superpotential ${\cal W}$. The K\"ahler potential ${\cal K}$ can be expressed as
\begin{equation}\label{K}
\kappa^2{\cal K}(\tau,\bar{\tau},S,\bar{S})=K(\tau,\bar{\tau},S,\bar{S}) - h \ln\left[-i(\tau-\bar{\tau})\right]\,,
\end{equation}
where $K(\tau,\bar{\tau},S,\bar{S})$ is the K\"ahler potential for the dilaton, the parameter $h$ is a dimensionless constant which depends on the choice of the number of compactified complex dimensions~\cite{Kobayashi:2016mzg}. The effective SUGRA description of the low-energy limit of superstring theory gives $h=3$~\footnote{In corresponding theory, the compactification of six dimensions will bring about three moduli $\tau_i\,(i=1,2,3)$ that corresponds to the radii of the three two-tori of the internal space and its standard form of K\"ahler potential~\cite{Cvetic:1991qm}
\begin{equation}
\label{eq:kahler2}{\cal K}=-\ln(S+\bar{S}) + \sum^{3}_{i=1}\ln\left[-i(\tau_i-\bar{\tau}_i)\right]\,,
\end{equation}
for which ${\cal K}$ is completely symmetric
under the exchange of the three $\tau_i$. We consider the minimal case, the symmetric point with $\tau_1=\tau_2=\tau_3=\tau$ that freezes all moduli fields except the single modulus $\tau$. This gives $h=3$.}. At the tree level, there is no $\tau$-dependence in the dilaton  K\"ahler potential, hence $K(\tau,\bar{\tau},S,\bar{S})=-\ln(S+\bar{S})$. When we consider the non-perturbative contributions, the corresponding dilaton K\"ahler potential is given by
\begin{equation}
K(\tau,\bar{\tau},S,\bar{S})=-\ln(S+\bar{S})+\delta k_{np}(S,\bar{S})\equiv K(S,\bar{S})\,,
\end{equation}
where the additional term $\delta k_{np}$ denotes the non-perturbative contribution from Shenker-like effects in heterotic string theories~\cite{Shenker:1990uf}. It provides non-perturbative corrections to K\"ahler potential within the $4D$ low-energy effective field theory, of the order of ${\cal O}(e^{-1/g^2_s})$, where $g^2_s$ represents the closed string coupling constant. A general parametrization of $\delta k_{np}$ is provided  by~\cite{Casas:1996zi,Higaki:2003jt}:
\begin{equation}
\delta k_{np}= d \left(\frac{S+\bar{S}}{2}\right)^{p/2} \exp\left(-b\sqrt{\frac{S+\bar{S}}{2}}\right)\,,
\end{equation}
where $p\,,b>0$ and $d$ is a real constant. The Shenker-term $\delta k_{np}$ is crucial to stabilize the dilaton and realize heterotic de Sitter vacua~\cite{Leedom:2022zdm}.

The combination $\tau-\bar{\tau}$ transform as $(-i\tau+i\bar{\tau})\rightarrow |c \tau + d|^{-2}(-i\tau+i\bar{\tau})$ under the action of modular symmetry. Therefore the moduli  K\"ahler potential transforms as follow under modular transformations
\begin{equation}
- 3 \ln\left[-i(\tau-\bar{\tau})\right] \rightarrow - 3 \ln\left[-i(\tau-\bar{\tau})\right] + 3 \ln(c \tau + d) + 3 \ln(c \bar{\tau} + d)\,.
\end{equation}
These extra terms has to be cancelled by the transformation of superpotential ${\cal W}$ to keep the modular invariance of the SUGRA function $\mathcal{G}$. This requires that the superpotential ${\cal W}$ be a modular function of weight $-h$ and its modular transformation is,
\begin{equation}
{\cal W} \rightarrow e^{i\delta(\gamma)}(c \tau + d)^{-h}{\cal W} \,,
\end{equation}
where $\delta(\gamma)$ is a phase depending on the modular transformation $\gamma$, and it is the so-called multiplier system.

The most general non-perturbative superpotential invariant under the modular symmetry~\cite{Cvetic:1991qm}: 
\begin{equation}
\label{eq:superpotential}{\cal W}(S,\tau)=\Lambda^3_W\frac{\Omega(S)H(\tau)}{\eta^6(\tau)}\,,
\end{equation}
where $\Lambda_W$ is the characteristic energy scale for this interaction. The function $\Omega(S)$ is technically arbitrary. It could take the form $\Omega(S)= c + e^{-S/b_a}$~\cite{Dine:1985rz}, which arises from gaugino condensation. Here $c$ is a constant and $b_a$ is related to the beta function of gauge group factor. We will assume the dilation field is stabilized as a premise. We leave a short discussion on dilation stabilization in Appendix~\ref{app:dilatonstablization}. The modulus field $\tau$ would serve as the inflaton. The modular function $H(\tau)$ is regular in the fundamental domain without singularities, and its most general form is~\cite{Cvetic:1991qm}: 
\begin{equation}
H(\tau)=(j(\tau)-1728)^{m/2}j(\tau)^{n/3}\mathcal{P}(j(\tau))\,,
\label{eq:H1}
\end{equation}
where $j(\tau)$ is the modular invariant $j$ function given in Eq.~\eqref{eq:jG4}, $\mathcal{P}(j(\tau))$ is an arbitrary polynomial function of $j(\tau)$, and both $m$ and $n$ are non-negative integers. To simplify the analysis, we choose the polynomial to be second order in $j(\tau)$:
\begin{equation}\label{eq:jpoly}
\mathcal{P}(j(\tau))=1+\beta\left(1-\frac{j(\tau)}{1728}\right)+\gamma\left(1-\frac{j(\tau)}{1728}\right)^2\,,
\end{equation}
where $\beta,\gamma$ are free real parameters. An equivalent parameterization of the $H$ function is given by~\cite{Cvetic:1991qm,Leedom:2022zdm}:
\begin{equation}
H(\tau) = \left(\frac{G_4(\tau)}{\eta^8(\tau)}\right)^{n}\left(\frac{G_6(\tau)}{\eta^{12}(\tau)}\right)^{m}\mathcal{P}(j(\tau))\,,
\label{eq:H2}
\end{equation}
where $G_4$ and $G_6$ are Eisenstein series of weight 4 and 6 respectively and their definition of $G_4$ and $G_6$ is given in Eq.~\eqref{eq:G2kE2k}. From Eqs.~\eqref{eq:jG4} and ~\eqref{eq:j-1728}, one can see that the two expressions of the $H$ function in Eq.~\eqref{eq:H1} and Eq.~\eqref{eq:H2} are equivalent up to a normalization constant. Notice that $H(\omega)=0$ for $n\geq 1$ and $H(i)=0$ for $m\geq 1$, as shown in Eq.~\eqref{eq:hfunc-mn}. 

The scalar potential in $\mathcal{N}=1$ supergravity is given by~\cite{Ibanez:2012zz},
\begin{equation}\label{VpG}
V(\tau,S)=e^{\kappa^2{\cal K}}({\cal K}^{\alpha\overline{\beta}}D_{\alpha}{\cal W}\overline{D_{\beta}{\cal W}} - 3\kappa^2|{\cal W}|^2)\,,
\end{equation}
where the covariant derivative is defined by $D_\alpha {\cal W} \equiv \partial_\alpha {\cal W} + \kappa^2{\cal W} (\partial_\alpha {\cal K})$ and ${\cal K}^{\alpha\overline{\beta}}$ is the inverse of the K\"ahler metric ${\cal K}_{\alpha\overline{\beta}}=\partial_\alpha\partial_{\overline{\beta}} {\cal K}$. The indices $\alpha$, $\beta$ run over all superfields for our discussion, $\alpha, \beta=\tau,S$. With the K\"ahler potential in Eqs.~(\ref{eq:kahler2},\ref{K}) and the superpotential in Eq.~\eqref{eq:superpotential}, we can straightforwardly obtain the scalar potential as follows:
\begin{equation}
V(\tau,S)=\Lambda^4  Z(\tau,\Bar{\tau})\left[(A(S,\Bar{S})-3)\abs{H(\tau)}^2+\hat{V}(\tau,\Bar{\tau})\right]\,,
\label{eq:scalarpotential}
\end{equation}
where we have defined $\Lambda=(\Lambda^6_W \kappa^2 e^{K(S,\bar{S})}\abs{\Omega(S)}^2)^{1/4} $ with
\begin{equation}\label{AVZ}
\begin{split}
A(S,\Bar{S})&=\frac{K^{S\bar{S}}D_S W D_{\Bar{S}}\bar{W}}{\abs{W}^2}=\frac{K^{S\bar{S}}\abs{\Omega_S+K_S\Omega}^2}{\abs{\Omega}^2}\,,\\
\hat{V}(\tau,\bar{\tau})&=\frac{-(\tau-\bar{\tau})^2}{3}\abs{H_\tau(\tau)-\frac{3i}{2\pi}H(\tau)\widehat{G}_2(\tau,\bar{\tau})}^2\,,\\
Z(\tau,\bar{\tau})&=\frac{1}{i(\tau-\bar{\tau})^3\abs{\eta(\tau)}^{12}}\,,
\end{split}
\end{equation}
where the subscript in $\Omega_S=\partial \Omega/\partial S$ and $H_\tau=\partial H /\partial \tau $ denotes the derivative with respect to the specific field. $\widehat{G}_2$ is the non-holomorphic modular form of weight 2, and its definition can be found in the Appendix~\ref{app:modularforms}. In addition, we see both the functions $\hat{V}$ and $Z$ are modular functions with weight 0. We have assumed that the dilaton sector is stabilized, thus $A(S,\bar{S})$ is treated as a free parameter in the model. The $A(S,\Bar{S})$ term is crucial for uplifting the potential. Once $A(S,\Bar{S})>3$, we can guarantee that the potential in Eq.~\eqref{eq:scalarpotential} is positive semi-definite.

%%%%%%%%%%%%%%%%%%%%%%%%%%%%%%%%%%%%%%%%%%%%%%%%%%%%%%%%%%%%%%%%%%%%%%%%%%
\subsection{The properties of the modular invariant scalar potential\label{sec:propertypotential}}

The vacuum structure of this potential at $\tau=i$ and at $\tau=\omega=e^{i 2\pi/3}$ has been extensively studied in~\cite{Leedom:2022zdm}, where they find the following results based on the choice of $(m,n)$ in Eq.~\eqref{eq:H1}:
\begin{itemize}
\item If $m=n=0$, then both fixed points can have a  de Sitter (dS) vacuum.
\item If $m>1,n=0$, then $\tau=\omega$ is a dS minimum, while $\tau=i$ is Minkowski minimum.
\item If $m=0,n>1$, then $\tau=i$ is a conditional dS minimum, which depends on the value of $A(S,\Bar{S})$. $\tau=\omega$ is always a Minkowski minimum.
\item If $m=1,n>0$ or $n=1,m>0$, the vacuum is unstable.
\item If $m>1,n>1$, then we always have Minkowski extrema in these two fixed points.
\end{itemize}

The scalar potential in Eq.~\eqref{eq:scalarpotential} is modular invariant. Consequently the derivatives $\partial_\tau V$ and $\partial_{\bar{\tau}} V$ are  weights (2,0) and (0,2) non-holomorphic modular functions respectively. Hence, they vanish at the fixed points:
\begin{equation}
\left.\frac{\partial V}{\partial \tau}\right|_{\tau=i,\omega} = \left.\frac{\partial V}{\partial \bar{\tau}}\right|_{\Bar{\tau}=i,\omega}=0\,.
\end{equation}
Moreover, the scalar potential is also invariant under $\tau\to -\bar{\tau}$. This can be proven by noticing the following transformation properties\footnote{We have assumed that the coefficients of the polynomial $\mathcal{P}(j)$ to be real.}:
\begin{equation}\label{eq:tautobartauFT}
\begin{split}
\eta(\tau) &\to \eta(\tau)^*\,, \quad  H(\tau) \to H(\tau)^*\,,\\
H_\tau  &\to -H_\tau^* \,,\quad \hat{G}_2 \to \hat{G}_2^* \,.
\end{split}
\end{equation}
This fact comes from the reality of the scalar potential. Together with the modular transformations $\tau\to \tau+1$ and $\tau \to -1/\tau$, they ensure the first derivative along certain directions at the boundary of the fundamental domain vanishes~\cite{Cvetic:1991qm}\footnote{\label{ft:deriv-zero}Let's write $\tau=x+i y$, where $x$ and $y$ are the real and imaginary parts of $\tau$, respectively. The combined modular symmetry $\tau\to \tau+1$ and reality condition $\tau\to -\Bar{\tau}$ tell us that the potential is invariant under $x\to -x-1,y\to y$ and $x=-1/2$ is a fixed point of the symmetry. Hence we have $V(x)=V(-x-1)$. Taking derivative with respect to $x$ on both sides yields:
\begin{eqnarray}
\nonumber
\left.\frac{\partial V(x)}{\partial x}\right|_{x=-1/2}=\left.\frac{\partial V(-x-1)}{\partial x}\right|_{x=-1/2}=-\left.\frac{\partial V(u)}{\partial u}\right|_{u=-1/2} \,,
\end{eqnarray}
where we have defined $u=-x-1$. Hence, the derivative vanishes at the fixed points $x=\pm 1/2$. For the second case, we can express $\tau=\rho e^{i\theta}$. The combined transformations of $\tau\to -1/\tau$ and $\tau\to -\Bar{\tau}$ indicate that the potential is invariant under $\rho\to 1/\rho,\theta\to \theta$. Hence $V(\rho)$=$V(1/\rho)$ and $\rho=1$ is a fixed point of this transformation, and consequently we get
\begin{equation}
\nonumber
\label{eq:partial-rho}  \left.\frac{\partial V(\rho)}{\partial \rho}\right|_{\rho=1}=\left.\frac{\partial V(1/\rho)}{\partial \rho}\right|_{\rho=1}=-u^2\left.\frac{\partial V(u)}{\partial u}\right|_{u=1}=-\left.\frac{\partial V(u)}{\partial u}\right|_{u=1} \,,
\end{equation}
with $u=1/\rho$. Thus the identiy $\left.\frac{\partial V(\rho)}{\partial \rho}\right|_{\rho=1}=0$ is satisfied.}:
\begin{equation}
\left. \frac{\partial V}{\partial Re(\tau)}\right|_{Re(\tau)=\pm 1/2,0} = 0 \,,\quad \left. \quad \frac{\partial V}{\partial \rho}\right|_{\rho=1}=0\,,
\end{equation}
where we have used $\tau=\textrm{Re}(\tau)+i \textrm{Im}(\tau)$ in the first equality and $\tau=\rho e^{i \theta}$ in the second equality. 
If ${\textrm{Re}}(\tau)$ or $\rho$ sits in a local minimum, we can neglect the motion of them at the corresponding boundaries. In this case, we have a single field inflation. One explicit example can be find in section~\ref{sec:MII}. Thus the modular symmetry provides a good reason to separate the modulus field to one inflaton field and another one perpendicular to inflation direction.

\begin{figure}[hbt]
\centering
\includegraphics[width=0.6\linewidth]{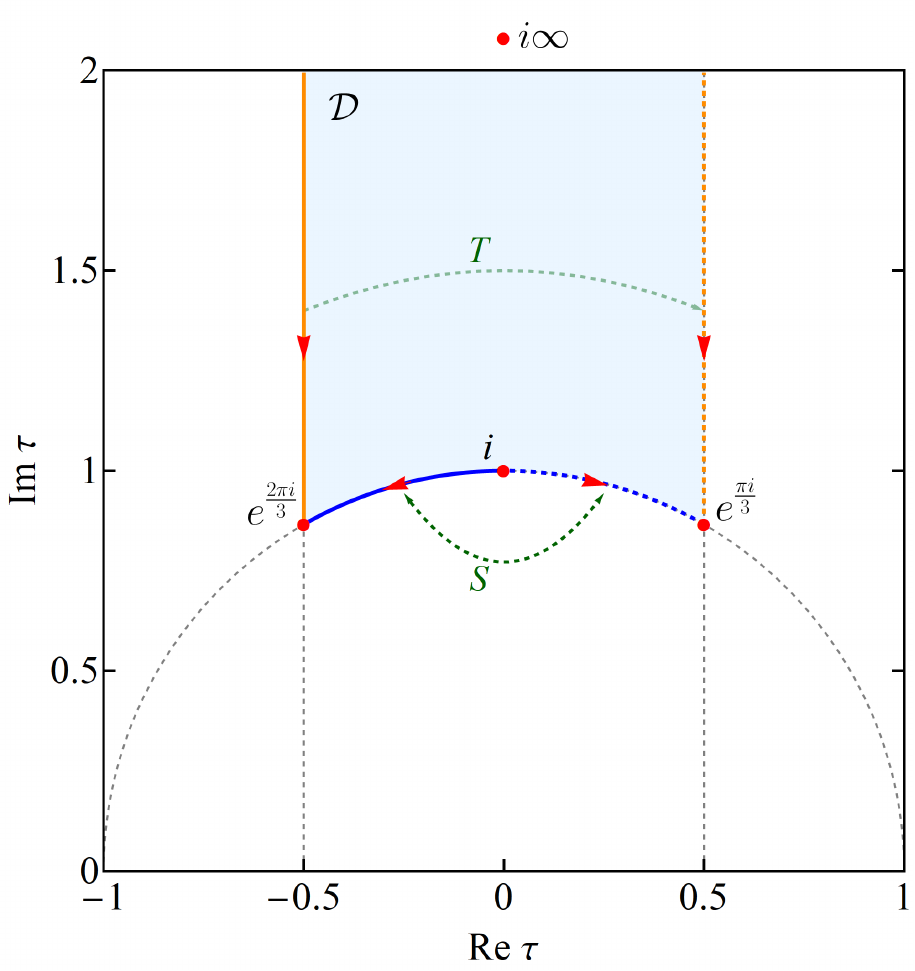}
\caption{The light blue region represents the fundamental domain $\mathcal{D}$ of the modular group, while the blue line denotes the inflationary trajectory from maxima $\tau=i$ to minima $\tau=\omega=e^{i 2\pi/3}$. Additionally, the blue dashed line depicts the inflaton slowly rolls from $i$ to $-\omega^2=e^{i \pi/3}$. Meanwhile, the orange line signifies the occurrence of accidental inflation to the point $\omega$.}
\label{fig:traj1}
\end{figure}

Motivated by the vacuum structure of modulus~\cite{Leedom:2022zdm} and the  special properties of fixed points $\tau=i,\omega=e^{i 2\pi/3}$ under modular transformation, we shall consider two different trajectories of inflation along the boundary of fundamental domain: 
\begin{itemize}
\item $m=0,n\geq2$, we consider slow roll along the lower boundary (arc) from one fixed point $i$ to another fixed point $\omega$.
\item $m\geq 2, n\geq 2$, we consider slow roll along the left boundary from $i \infty $ to the fixed point $\omega$.
\end{itemize}
We illustrate these two trajectories in figure~\ref{fig:traj1}. We will show some concrete examples below where the scalar potential is flat enough to accommodate inflation.

%%%%%%%%%%%%%%%%%%%%%%%%%%%%%%%%%%%%%%%%%%%%%%%%%%%%%%%%%%%%%%%%%%%%%%%%%%
\subsection{Constraints on the scalar potential from inflation and modular stabilization}

Let's first focus on the case where inflation occurs at the lower boundary. The modular symmetry fixed points $\tau=i,\omega$ play a special role in moduli stabilization, where they are separately considered to be a minimum~\cite{Cvetic:1991qm,Leedom:2022zdm}. For inflation, we would like to have one of them to be the minimum of the scalar potential and another one to be the saddle point. We are interested in how they are connected. The saddle point of the potential could serve as the starting point for inflation. The minimum of the potential has to be stable, hence the possibility of $m=1$ or $n=1$ is excluded. The whole potential has to be non-negative in the fundamental domain, consequently we always require $A(S,\bar{S})\geq3$.

In the following, we explore the constraints on the parameters of the scalar potential from modular inflation. As mentioned earlier, in order to realize the slow roll inflation drived by $\tau$, we would like one of the fixed points to be a saddle point and the other to be the minimum. This can not be achieved for $m\geq 2$ and $n=0$, as both fixed points are local minima of the scalar potential. In principle, there exists a parameter space for $m=0$ and $n=0$ where $\tau=i$ is a saddle point and $\tau=\omega$ is a dS minimum. We may refer to $V(i)$ as the inflation scale and $V(\omega)\approx 10^{-122}$ as the cosmological constant today. However, the ratio of their value reads:
\begin{equation}
\frac{V(i)}{V(\omega)}=\frac{\Gamma^{18}(1/3)}{\Gamma^{12}(1/4)}\frac{\abs{\mathcal{P}(1728)}^2}{\abs{\mathcal{P}(0)}^2}\,.
\end{equation}
Thus for a TeV scale inflation where $V(i)\approx 10^{-30}$, we need $\abs{\mathcal{P}(1728)}^2/\abs{\mathcal{P}(0)}^2\approx 10^{90}$. Given that $j(\tau)$ varies by only $10^3$ from $\tau=i$ to $\tau=\omega$, it seems very unnatural to consider such a huge difference in the polynomial $\mathcal{P}(j)$. Hence we do not consider the choice $m=n=0$ in this work~\footnote{The modular inflation along the unit arc for $m=n=0$ is studied in Ref.~\cite{King:2024ssx} which appeared in arXiv after our submission.}.

\begin{figure}[hbt!]
\centering    \includegraphics[width=.90\linewidth]{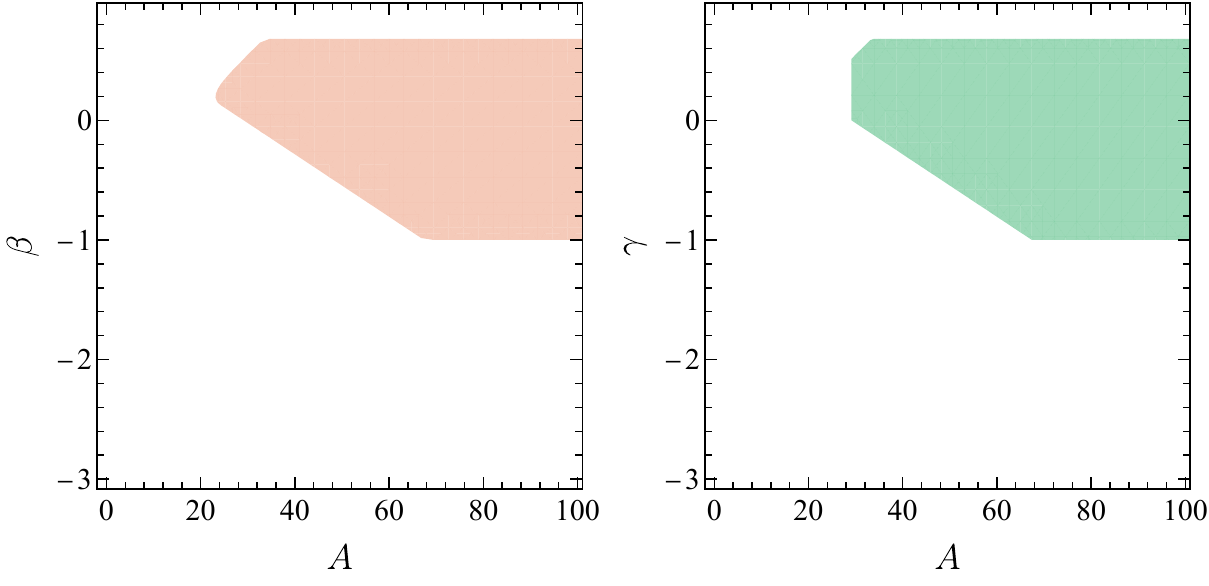}
\caption{In the left panel, we examine the parameter space $(A, \beta)$ with $\gamma=0$, constrained by the condition $\partial V/\partial \theta < 0$. Similarly, the right panel explores the parameter space $(A, \gamma)$ with $\beta=0$. Here, we illustrate this concept using the example $n=2$.}
\label{fig:ImtauD1}
\end{figure}

Another possibility is $n\geq 2$ and $m=0$ where $\tau=i$ is a saddle point and $\tau=\omega$ is a global minimum with $V(\omega)=0$. To explicitly discuss this possibility, we should calculate the Hessian matrix at the fix points. The first derivative of the scalar potential vanishes and the second derivatives at $\tau=i$ read:
\begin{equation}\label{eq:Vm0taui}
\begin{split}
\partial^2_\tau V\Big|_{\tau=i} &=  -{\cal C}_n (A-1) {\cal B}_n\,,\quad  \partial_\tau\partial_{\bar{\tau}}V\Big|_{\tau=i} ={\cal C}_n(A-2+ |{\cal B}_n|^2)\,.
\end{split}
\end{equation}
where the functions ${\cal B}_n$ and ${\cal C}_n$ are:
\begin{eqnarray}
\nonumber{\cal B}_n &\equiv& \frac{\Gamma^{8}(1/4)}{192\pi^4}\left(1+8n+41472 \frac{ {\cal P}'(1728)}{ {\cal P}(1728)} \right)\,,\\
\label{eq:calBnCn} {\cal C}_n &\equiv& \frac{(2\pi)^94^{2n-1}3^{2n+1}}{\Gamma^{12}(1/4)}\left|{\cal P}(1728)\right|^2\,,
\end{eqnarray}
where $\mathcal{P}'(j)\equiv d\mathcal{P}/dj$ is the derivative of the polynomial function $\mathcal{P}(j)$ with respect to $j$. For inflation along the unit arc in the complex $\tau$ plane, it is convenient to calculate the Hessian matrix in the polar coordinates $\tau=\rho e^{i\theta}$, and the elements of the Hessian matrix at the starting point $\tau=i$ are determined to be:
\begin{equation}\label{eq:p2Vpolar}
\begin{split}
\frac{\partial^2 V}{\partial \rho^2}&= 2{\cal C}_n\left[A-2+ |{\cal B}_n|^2+ (A-1)\text{Re}({\cal B}_n)\right]\,,\\
\frac{\partial^2 V}{\partial \theta^2}&=  2{\cal C}_n\left[A-2+ |{\cal B}_n|^2 - (A-1)\text{Re}({\cal B}_n)\right]\,,\\
\frac{\partial^2 V}{\partial \theta~\partial \rho} &= -2{\cal C}_n (A-1) \text{Im}({\cal B}_n)\,.
\end{split}
\end{equation}
From Eq.~\eqref{eq:calBnCn}, we see that ${\cal C}_n$ is a positive number and the imaginary part of ${\cal B}_n$ vanishes at $\tau=i$, since the coefficients of the polynomial ${\cal P}(j)$ are real in our setting and thus ${\cal P}$ is real at $\tau=i$. Consequently the cross derivative $\frac{\partial^2 V}{\partial \theta~\partial \rho}$ is exactly zero at $\tau=i$. In fact, this cross derivative is always vanishing anywhere at the boundary $|\tau|=1$ of the fundamental domain, as can be seen from the footnote~\ref{ft:deriv-zero}. For the inflaton $\tau$ rolling along the unit arc,  the scalar potential at the starting point $\tau=i$ should reach a maximum along angular direction $\theta$ while it gets a minimum along the radial direction $\rho$. As a consequence, the following conditions should be satisfied~\footnote{The scalar potential at $\tau=i$ is
\begin{eqnarray}
V(i) =\Lambda^4 (A-3) 12^{2n}\frac{(2\pi)^9}{\Gamma^{12}(1/4)}\left|{\cal P}(1728)\right|^2\,.
\end{eqnarray}}
\begin{equation}
\frac{\partial^2 V}{\partial \rho^2}\bigg|_{\tau=i}>0\,, \quad \frac{\partial^2 V}{\partial \theta^2}\bigg|_{\tau=i}<0\,, \quad V(i)>0,
\end{equation}
which leads to
\begin{equation}\label{eq:rangeAm0}
A>2+{\cal B}_n \,,\quad\quad  {\cal B}_n >1    \,.
\end{equation}
The condition ${\cal B}_n >1$ restricts the polynomial parameter in Eq.~\eqref{eq:jpoly} as follow,
\begin{equation}
\beta < \frac{1}{24}\left(8n+1-\frac{192\pi^4}{\Gamma^{8}(1/4)}\right)\,.
\end{equation}
Moreove, another condition $A>2+{\cal B}_n$ can be written into a more suggestive form:
\begin{equation}\label{eq:Aregforpoly}
A>\frac{\Gamma^{8}(1/4)}{192\pi^4}\left(1+8n - 24\beta\right) + 2\,.
\end{equation}
In particular, when $\beta =0$ and $n=2$, the parameter $A$ should be greater than $29.132$. Besides the local properties of the scalar potential at the fix points, we would also like to impose an additional constraint along the inflationary trajectory $\pi/2 < \theta < 2\pi/3 $:
\begin{equation}
\frac{\partial V}{\partial \theta}\bigg|_{\rho=1}<0\,,\quad  \frac{\partial^2 V}{\partial \rho^2}\bigg|_{\rho=1}>0\,,
\end{equation}
where the first condition ensures that the inflaton smoothly rolls down to the minimum at $\tau=\omega$, and the second one ensures $\rho=1$ is a local minimum along the inflation trajectories. The algebraic expression for this condition is rather complicated and we solve it numerically. The first constraint is demonstrated in figure~\ref{fig:ImtauD1} and the second one is verified for each inflation potentials.

The potential at the fixed point $\tau=\omega$ or $\tau=-\omega^2$ is much simpler, we have:
\begin{equation}
V (\omega)= 0\,,\quad \partial^2_\tau V(\omega) =0\,,\quad \partial_\tau\partial_{\bar{\tau}} V(\omega)\geq 0     \,.
\end{equation}
From Eq.~\eqref{eq:scalarpotential}, as long as $A\geq3$ the potential is non-negative in the whole complex space. Hence, $V (\omega)= 0$ is enough to ensure $\tau=\omega$ is a global minimum of the scalar potential.

%%%%%%%%%%%%%%%%%%%%%%%%%%%%%%%%%%%%%%%%%%%%%%%%%%%%%%%%%%%%%%%%%%%%%%%%%%
\section{Modular invariant inflation\label{sec:MII}}

In this section, we shall show that the scalar potential in Eq.~\eqref{eq:scalarpotential}, which has been used to address the modulus stabilization~\cite{Cvetic:1991qm,Leedom:2022zdm}, can also naturally realize the slow roll inflation. We are concerned with the inflationary trajectories along the boundary of fundamental domain in the following.

\subsection{Slow roll along the unit arc}
In this scenario, it might be useful to rewrite the scalar potential in terms of the radial and angular components, $\tau=\rho e^{i \theta}$. The kinetic term of the modulus reads as,
\begin{equation}
\mathcal{L}_{\text{kin}}=  \frac{\partial^2\mathcal{K}}{\partial \tau \partial \bar{\tau}} \partial_{\mu} \tau \partial^{\mu} \bar{\tau}=\frac{3}{(-i\tau+i\bar{\tau})^2}\partial_{\mu} \tau \partial^{\mu} \bar{\tau}=\frac{3}{4 \sin^2\theta}\left( \frac{1}{\rho^2}\partial_{\mu} \rho \partial^{\mu} \rho+\partial_{\mu} \theta \partial^{\mu} \theta \right)\,.
\end{equation}
As shown in Eq.~\eqref{eq:partial-rho}, the modular invariance of the scalar potential requires $\partial V /\partial \rho|_{\rho=1}=0$. Here and hereafter, we will always set $\rho=1$ and keep $\theta$ as the only degree of freedom. To normalize the kinetic term of $\theta$, we further introduce the canonical field $\phi=\sqrt{3/2}\ln(\tan(\theta/2))$. As an example, we have $\phi=0$ when $\theta=\pi/2$.

In this section, we consider the case where $m=0,n\geq2$ in Eq.~\eqref{eq:H1}. This potential has a local maximum at $\tau=i$ and a local Minkowski minimum at $\tau=\omega$. As mentioned earlier, the modular symmetry ensures that the first derivative of the potential vanishes at $\tau=i$, which motivates us to investigate the inflation near this point. The inflation trajectory is shown in figure~\ref{fig:traj1}. The inflation phenomenology can be approximated by its Taylor expansion near $\tau=i\,(\phi=0)$. The full potential in Eq.~\eqref{eq:scalarpotential} can be approximated by the following term during inflation:
\begin{equation}
V(\phi)=V_0(1-\sum_{k=1}^{\infty}C_{2k}\phi^{2k})\,,
\end{equation}
where each coefficient depends on the choice of $A(S,\bar{S}),(m,n)$ and the parameterzation of the polynomial function $\mathcal{P}(j)$. Note that the potential is an even function of $\phi$, which arises from the ${\cal S}$ symmetry of the modular group. Along the arc, the ${\cal S}$ symmetry $\tau \to -1/\tau$ indicates a $Z_2$ symmetry in terms of the canonically normalised field $\phi \to -\phi$. We will focus on the case where $\phi>0$. During inflation, we find the potential is mostly dominated by the terms $C_2 \phi^2$ and $C_{2p}\phi^{2p}$, where $p$ is a specific integer. This implies that $0<\abs{C_{2p'}}\ll C_{2p}$ for all $p'<p$. Let's first investigate this simplified potential:
\begin{equation}
V(\phi)=V_0(1-C_2\phi^2-C_{2p}\phi^{2p})\,,
\end{equation}
The slow-roll parameters read:
\begin{equation}\label{eq:SRparas}
\begin{split}
\varepsilon_V&=\frac{1}{2}\left(\frac{V'}{V}\right)^2=\frac{1}{2}\left(\frac{-2C_2\phi-2pC_{2p}\phi^{2p-1}}{1-C_2\phi^2-C_{2p} \phi^{2p}}\right)^2\,,\\
\eta_V&=\frac{V''}{V}=\frac{-2C_2-2p(2p-1)C_{2p}\phi^{2p-2}}{1-C_2\phi^2-C_{2p} \phi^{2p}}\,,
\end{split}
\end{equation}
where $C_2\ll C_{2p}$ and $ C_{2p} \gg 1$. We are interested in the region where $\phi \ll 1$, which means $\varepsilon_V \approx (\eta_V \phi)^2 \ll \eta_V$. Note the starting point of the observable inflation, $\phi_*$, can be calculated from:
\begin{equation}
n_s = 1-6\epsilon(\phi_*)+2\eta(\phi_*) \approx 1+2\eta(\phi_*) \approx 1- 4C_2 - \mathcal{O}(\phi_*^2)\,,
\end{equation}
and the CMB observation suggests $n_s \approx 0.9649 $~\cite{Planck:2018jri}. For our setup, we would require $1-4C_2>n_s$; otherwise, the spectral index will be smaller than the observed value of CMB. This approximately implies $0< C_2 <0.008$.

The starting point $\phi_*$ and end point $\phi_e$ of observable inflation are controlled by $\eta$ and corresponding field values are given by,
\begin{equation}
\begin{split}
\phi_* &\approx \left(-\frac{\eta_*+2 C_2}{ 2p (2p-1)C_{2p}}\right)^{\frac{1}{2p-2}}\,,\\
\phi_e &\approx \left(-\frac{\eta_e+2 C_2}{ 2p (2p-1)C_{2p}}\right)^{\frac{1}{2p-2}}\,,
\end{split}
\label{eq:phivalue}
\end{equation}
where we have defined $\eta_*=(n_s-1)/2$, $\eta_e=-1$, and used approximation $1-C_2\phi^2-C_{2p} \phi^{2p}\approx 1$. This can be verified by substituting the above solution into the expression, which is suppressed by large $C_{2p}$. The number of e-folds is given by,
\begin{equation}
\begin{split}
N_e =&\int_{\phi_*}^{\phi_e}\frac{1}{\sqrt{2\varepsilon_V(\phi)}}d\phi\\\approx& \int_{\phi_*}^{\phi_e} \frac{1}{2C_2\phi+2pC_{2p}\phi^{2p-1}}d\phi\\ =& \left. \frac{\ln{( 2p C_{2p}+2 C_2 \phi^{2-2p})}}{2C_2 (2p-2)}\right|_{\phi_*}^{\phi_e} \\=&\frac{1}{2C_2(2p-2)}\left(\ln{\frac{\eta_e+(4-4p)C_2}{\eta_e+2C_2}}-\ln{\frac{\eta_*+(4-4p)C_2}{\eta_*+2C_2}}\right)\,,
\end{split}
\label{eq:Ne}
\end{equation}
which initially decrease with $C_2$ and then increases with it in the region where $0<C_2<0.008$. The minimum of e-folds,  $N_{e,\textrm{min}}$, is approximately $77$ for $n_s=0.9649$ and $p=2$, and approximately $50$ for  $n_s=0.9649$ and $p=3$. Thus $p=2$ always generates too many e-folds, while $p\geq3$ could be consistent with our current observations. It is therefore necessary to ensure $C_2$ and $C_4$ are much smaller than $C_6$ in the expansion.

We will address the coefficients $C_{2n}$ in the following section. To do this, we need the local expansion of the $j$-invariant and $\eta$ functions in terms of the canonically normalized field $\phi$:
\begin{equation}
\begin{split}
j(\phi) &\approx 1728(1 - 9.579 \phi^2+ 40.142\phi^4-102.618\phi^6)+\mathcal{O}\left(\phi^8\right) \,,\\
\abs{\eta(\phi)}&\approx 0.768+0.083\phi^2-0.028\phi^4+0.005\phi^6+\mathcal{O}\left(\phi^8\right)\,,
\end{split}
\label{eq:jexpansion}
\end{equation}
and the explicit form of the polynomial $\mathcal{P}(j)$. For illustration, we will consider the quadratic form of $\mathcal{P}(j)$ parameterized as:
\begin{equation}\label{eq:polynimal1}
\mathcal{P}\left(j(\tau)\right)=1 + \beta\left(1-\frac{j}{1728}\right)+ \gamma\left(1-\frac{j}{1728}\right)^2 + \ldots\,.
\end{equation}
As one can see from Eq.~\eqref{eq:jexpansion}, the parameter $\beta$ controls the $\phi^2$ term, while $\gamma$ governs the $\phi^4$ term in the polynomial $\mathcal{P}(j)$. The overall factor $V_0$ of the scalar potential is determined to be
\begin{equation}
V_0=\frac{12^{2n}(2\pi)^9}{\Gamma^{12}(1/4)}\Lambda^4\left[A(S,\bar{S})-3 \right]\,.
\end{equation}

\subsubsection{$\mathcal{P}(j)=1$}
\begin{figure}[hbt]
\centering
\begin{subfigure}[b]{0.45\textwidth}
\centering
\includegraphics[width=\textwidth]{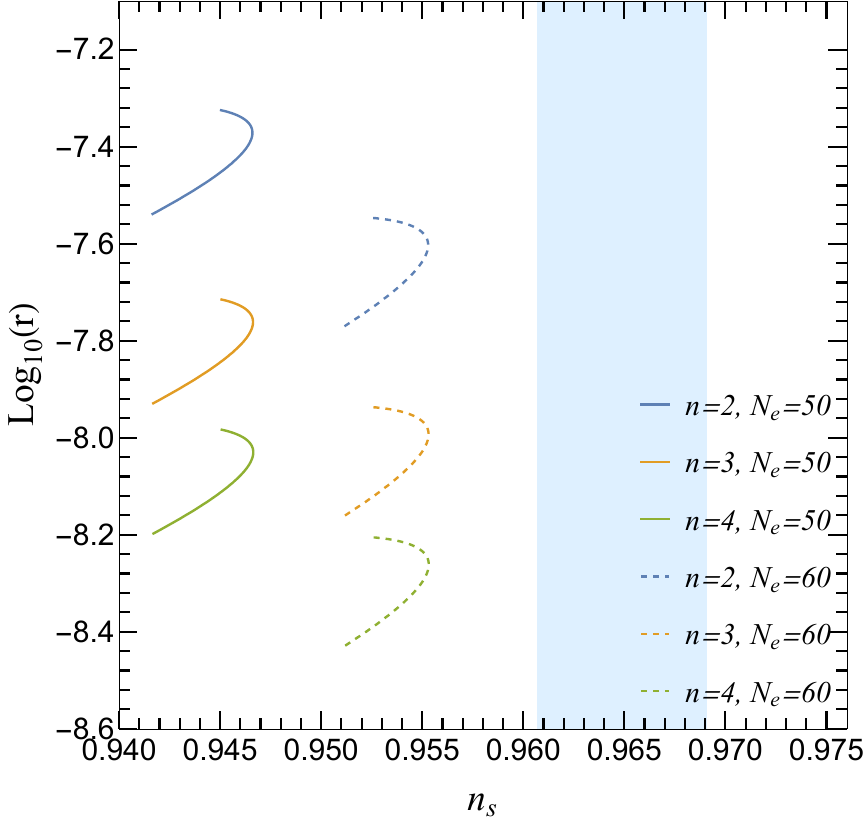}
\caption{$\mathcal{P}(j)=1$.}
\end{subfigure}
\hfill
\begin{subfigure}[b]{0.45\textwidth}
\centering
\includegraphics[width=\textwidth]{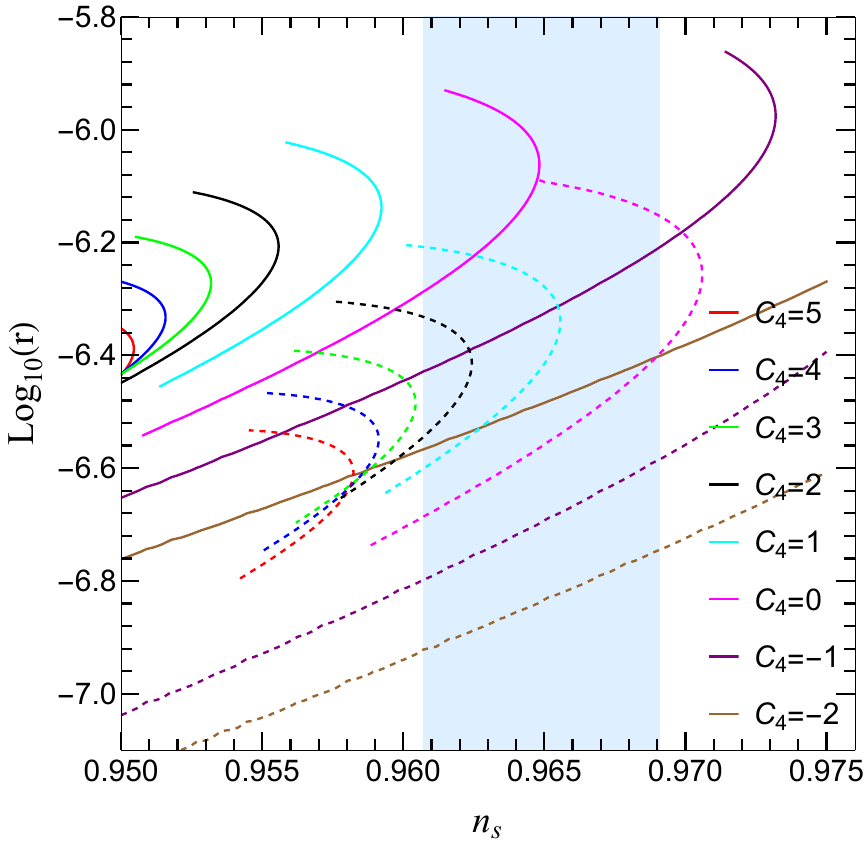}
\caption{ $\mathcal{P}(j)=1+\beta(1-j/1728)$.}
\end{subfigure}
\par\bigskip
\begin{subfigure}[b]{0.45\textwidth}
\centering
\includegraphics[width=\textwidth]{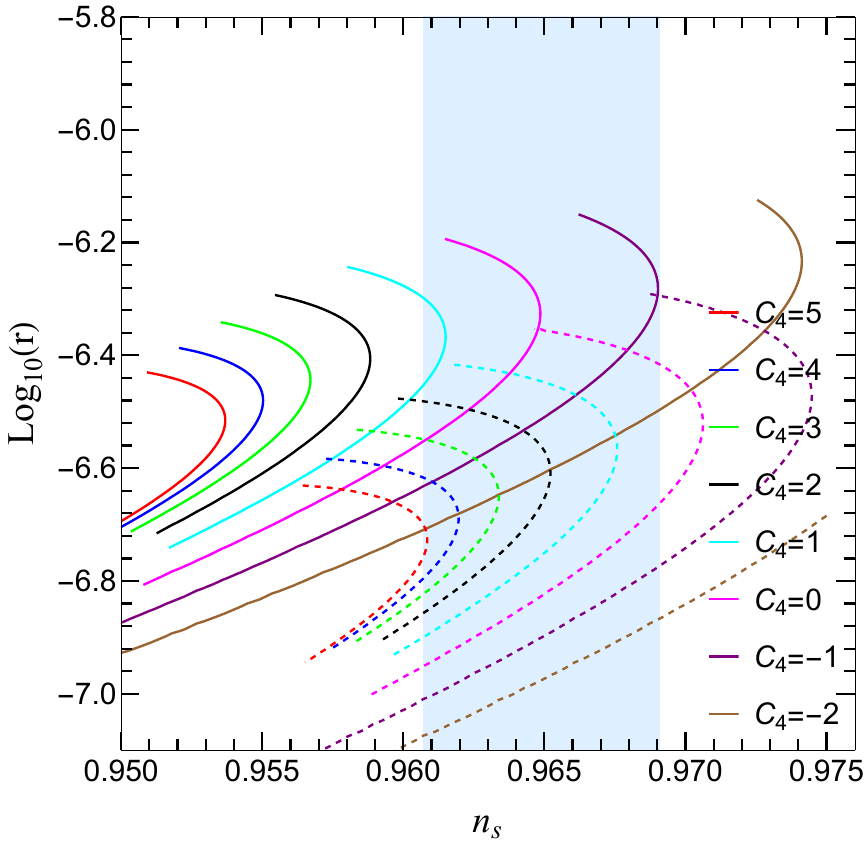}
\caption{ $\mathcal{P}(j)=1+\gamma\left(1-j/1728\right)^2$.}
\end{subfigure}
\hfill
\begin{subfigure}[b]{0.45\textwidth}
\centering
\includegraphics[width=\textwidth]{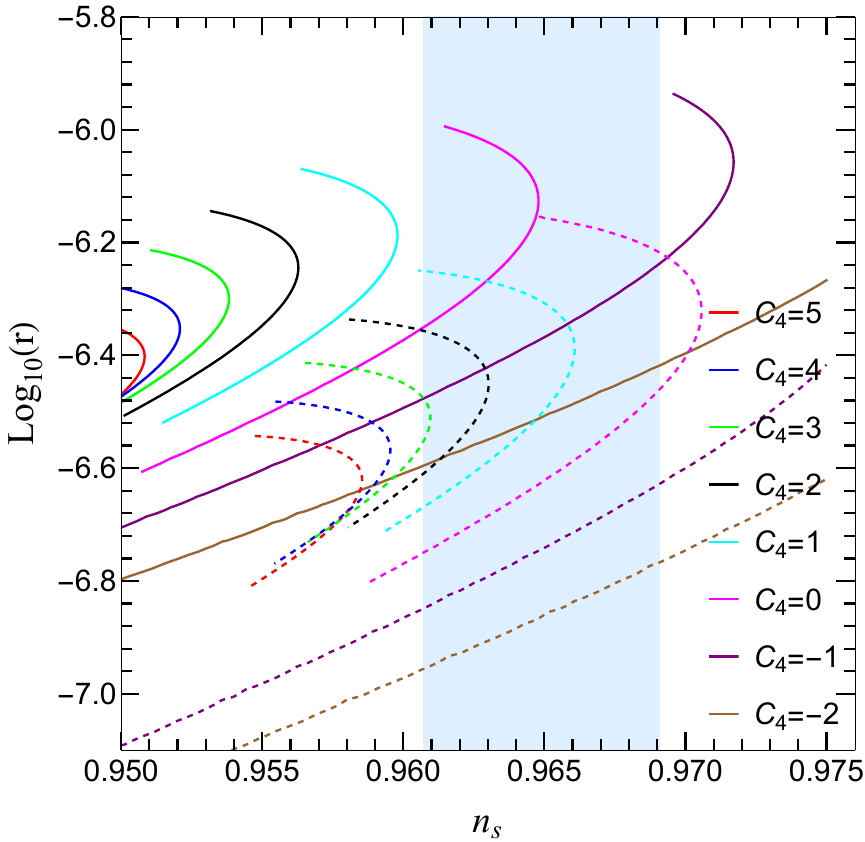}
\caption{$\mathcal{P}(j)=1+\beta(1-j/1728)+\gamma\left(1-j/1728\right)^2$.}
\end{subfigure}
\caption{\label{fig:inflationprediction} We present theoretical predictions of modular invariant inflation with different choices of the polynomial $\mathcal{P}(j)$. When $\mathcal{P}(j) = 1$,  the parameter $A$ varies within the region $3.596 + 12.771 n < A < 3.612 + 12.771 n$ given by Eq.~\eqref{eq:Avalue}. Conversely, when $\mathcal{P}(j)\neq 1$, we fix $n=2$ for plotting purposes. Additionally, we select $C_2$ and $C_4$ as the physical parameters and plot the lines by varying $C_2$ while holding $C_4$ constant. In the last panel, we set $A(S,\Bar{S})=25$. The $x$-axis represents the spectral index of the CMB power spectrum $n_S$, while the $y$ axis is the tensor-to-scalar ratio $r$ on a logarithm scale. Solid lines indicate predictions for $N_e=50$, whereas dashed lines are the predictions for $N_e=60$. Different colors denotes varying choices of $C_4$.}
\end{figure}

We start our study by considering the simplest case, $P(j)=1$. In this scenario, the shape of the potential is entirely determined by the dilaton contribution $A(S,\Bar{S})$. The potential can then be approximated by:
\begin{equation}
\begin{split}
C_2 =& 0.298+6.386 n -\frac{0.178+(7.617+81.554n)n}{A-3}\,,\\
C_4 =& -0.438+1.917n-20.389n^2+\frac{0.992+14.387n-122.250n^2+520.779n^3}{A-3}\,,\\
C_6 =& 0.234+9.024n-18.321n^2+43.398n^3\\&+\frac{-1.998-23.195n-718.051n^2+1247.593n^3-1662.767n^4}{A-3}\,.
\end{split}
\label{eq:C2bj0}
\end{equation}
The requirement for $0<C_2<0.008$ imposes an algebraic constraint on $A$:
\begin{equation}
3.596 + 12.771 n < A < 3.612 + 12.771 n\,,
\label{eq:Avalue}
\end{equation}
which suggests that $A$ is tightly constrained. In this case, the corresponding $C_4$ and $C_6$ are given by:
\begin{equation}
\begin{split}
C_4 &\approx 1.228+n(-9.571 + 20.389 n)\,,\\
C_6 &\approx 1.289-\frac{0.206}{0.047 + n} +  n \left[-52.076+ (85.482 - 86.797 n) n\right]\,.
\end{split}
\end{equation}
For $n=2$, this roughly means $A \approx 29.142$, $C_4 \approx 63.640$ and $C_6 \approx-455.408$. One can extend these results easily for general $n$. Even though $\abs{C_6}>C_4$, we have verified that the $\phi^4$ contribution is larger than the $\phi^6$ contribution during inflation, and the end of the inflation is controlled by the $\phi^4$ term. It is also interesting to consider the cases with large $n$. Note $C_4$ scales as $n^2$ while $C_6$ scales as $n^3$, which might raise concerns that the later term will eventually outgrows the former. However, this is not the case. We can examine their relative contributions at the end of inflation, where the $C_6\phi^6$ contributes the most. Using Eq.~\eqref{eq:phivalue} with $p=2$ and maintaining the assumption that $C_4\phi^4$ is much larger than $C_6\phi^6$, we have:  
\begin{equation}
\phi_e^2= \frac{1}{12C_4}\,,\quad\quad
\frac{C_6\phi^6}{C_4\phi^4}=\frac{C_6}{12C_4^2} \varpropto  \frac{1}{n}\,,
\end{equation}
which tells us that the relative contribution of the $C_6\phi^6$ term is suppressed by large $n$. These results perfectly align with our assumption, confirming that higher-order terms never dominate the potential. Meanwhile, the relative large coefficient $C_4$ leads to an overproduction of the number of e-folds if we choose $n_s$ to be around its central value. i.e. To achieve $50<N_e<60$, $n_s$ must be smaller than the CMB measurement. Theoretical prediction for this case can be found in figure~\ref{fig:inflationprediction}. We use solid and dashed lines to represent the results for $N_e=50$ and $N_e=60$, respectively, with different colors labeling different choices of $n$. Increasing $n$ does not change the preferred region of the spectral index but does decrease the tensor to scalar ratio. However, the spectral index lies between $0.942$ and $0.955$, which is excluded by the Planck 2018 results. Thus, we conclude that while the simplest choice of $\mathcal{P}(j)$ can support inflation, it does not fit our observational constraints within this framework.
%%%%%%%%%%%%%%%%%%%%%%%%%%%%%%%%%%%%%%%%%%%%%%%%%%%%%%%
%%%%%%%%%%%%%%%%%%%%%%%%%%%%%%%%%%%%%%%%%%%%%%%%%%%%%%%%%%%%%%
\subsubsection{$\mathcal{P}(j)=1 + \beta\left(1-j/1728\right)$}
\begin{figure}[thb]
\centering
\includegraphics[width=\textwidth]{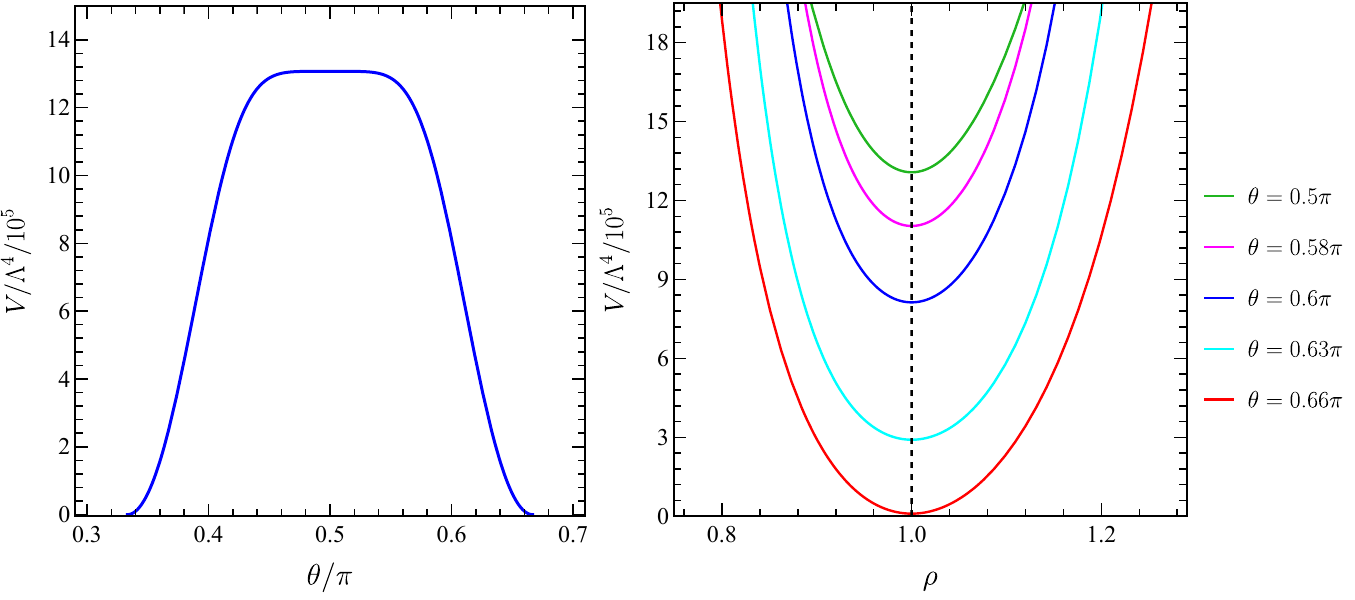}
\caption{When the potential has parameters $m=0$, $n=2$, $A=24.3091$, and $\beta=0.126425$, the left panel demonstrates the cross-section of the scalar potential via $\rho=1$ during inflation. Similarly, the right panel displays the cross-sections of scalar potential via various $\theta$ values throughout the inflationary process.}
\label{fig:model1-potential}
\end{figure}

\begin{figure}
\centering
\begin{subfigure}[b]{\textwidth}
\centering
\includegraphics[width=0.97\textwidth]{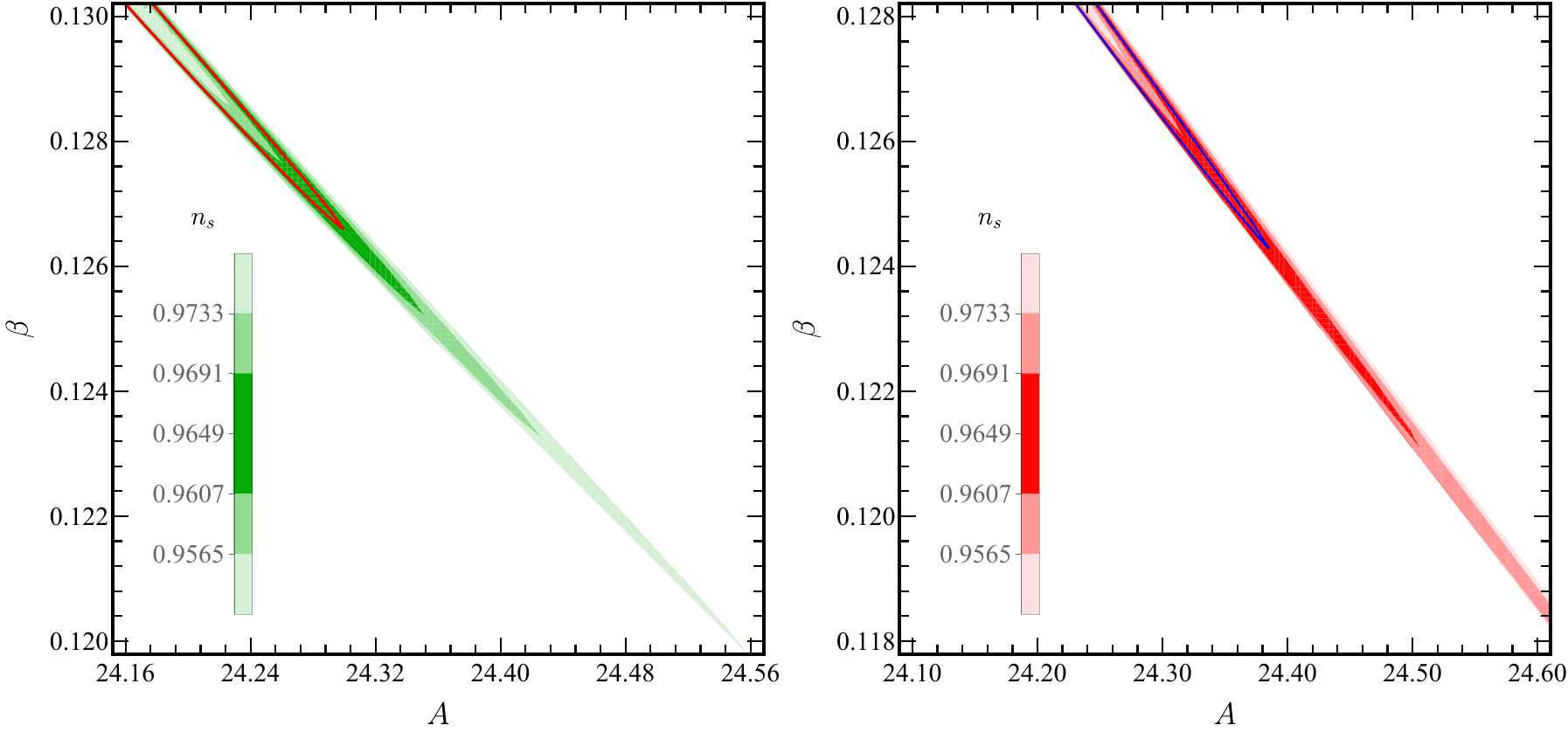}
\caption{$\mathcal{P}(j)=1+\beta(1-j/1728)$\,,}
\end{subfigure}
\par\bigskip
\begin{subfigure}[b]{1\textwidth}
\centering
\includegraphics[width=0.99\textwidth]{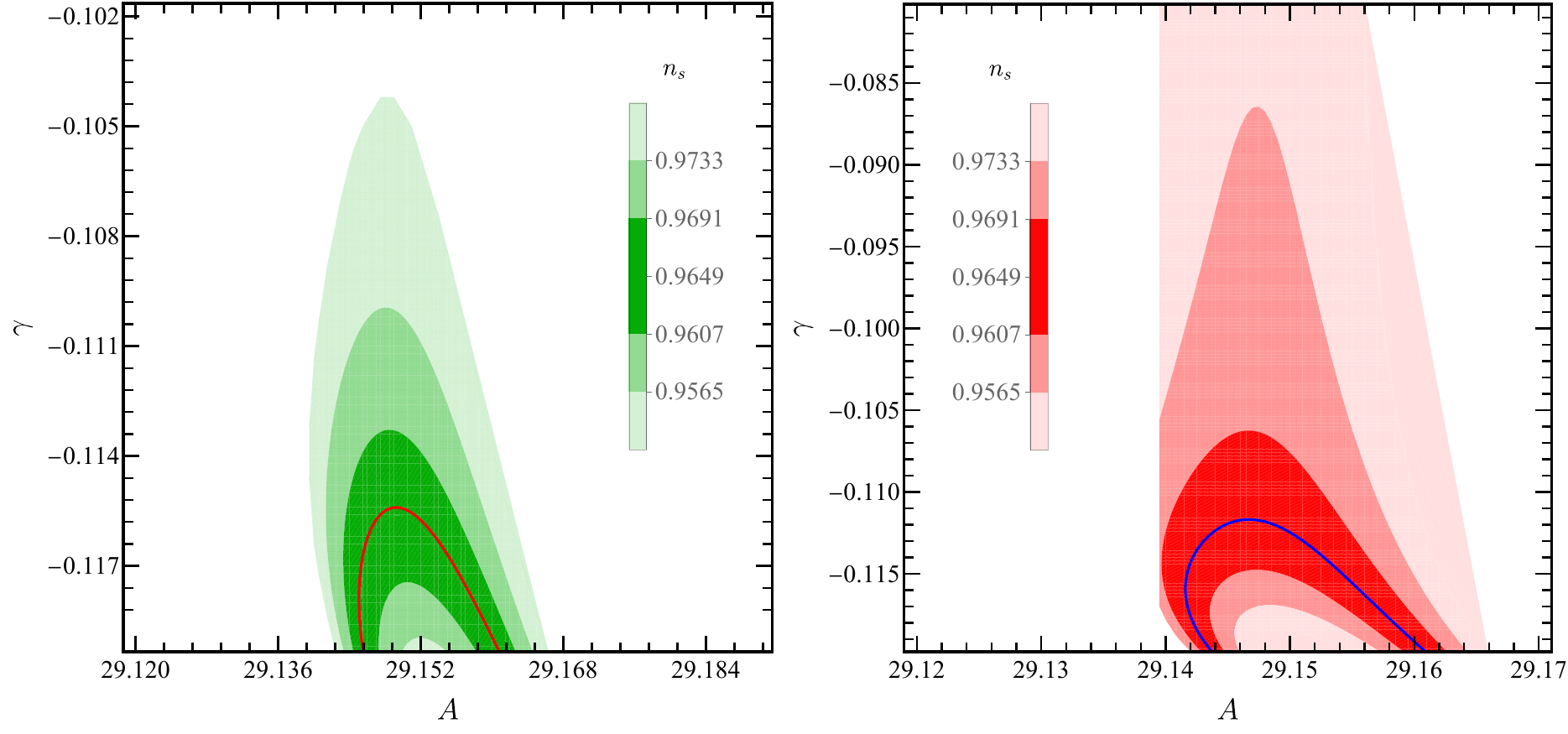}
\caption{$\mathcal{P}(j)=1+\gamma\left(1-j/1728\right)^2$.}
\end{subfigure}
\caption{These panels illustrate contour plot of the spectral index $n_s$ across the parameter planes $(A,\beta)$(upper sector) and $(A,\gamma)$(lower sector). In the left segment, the deep green region displays the $68\%$ CL region, with the red line indicating the contour of central value of $n_s$ for $50$ e-folds. The right segment demonstrates the distribution of $n_s$, with the deep red region and blue line representing $68\%$ CL region and its isopleth of central value for $60$ e-folds, as documented in   ~\cite{Planck:2018jri}. Furthermore,  it's important to note that these panels adhere entirely to the constraint specified in Eq.~\eqref{eq:Aregforpoly}.}
\label{fig:parameterspace}
\end{figure}

The next simplest choice would be $\beta \neq 0$. In this case, we have two free parameters: $A$ and $\beta$, with two algebraic constraints: $ 0<C_2<0.008 $ and $0<C_4\ll C_6$. The additional $\beta$ dependent terms in $C_2,C_4,C_6$ are given by:
\begin{equation}
\begin{split}
C_2 =&C_2|_{\beta=0}+C_{2,\beta}\\=& C_2|_{\beta=0} -19.157\beta+\frac{-733.987\beta^2+22.851\beta+489.325\beta n}{A-3}\,,\\
C_4=&C_4|_{\beta=0}+C_{4,\beta}\\=&C_4|_{\beta=0}-91.748\beta^2+\beta(85.997+122.331n)\\&+\frac{\beta^2(12741.904+9374.025n)+\beta(-262.043-3953.512n-4687.013n^2)}{A-3}\,,\\
C_6=&C_6|_{\beta=0}+C_{6,\beta}\\=&C_6|_{\beta=0}+\beta^2(796.369+585.877n)-\beta(237.566+475.949n+390.584n^2)\\&+\frac{\beta^2(106147.206+133975.980n+52377.172n^2)}{A-3}\\
&-\frac{\beta(1383.980+16226.188n+26183.928n^2+19953.208n^3)}{A-3}\,,
\end{split}
\end{equation}
where we have separated the coefficients into $\beta$ dependent and independent terms. The $C_{2,4,6}|_{\beta=0}$ are the same as those given in Eqs.~\eqref{eq:C2bj0}. The solution for $ 0<C_2<0.008 $ is:
\begin{equation}
3.596-38.314 \beta+12.771 n < A < 3.612-38.314 \beta+12.771 n\,.
\end{equation}
We further choose $-2<C_4<5$, such that $\abs{C_4}\ll C_6$. If $C_4\gg1$, the inflation potential will be dominated by the $\phi^4$ term. We have argued such a case is ruled out by the small spectral index $n_s$. When $C_4<-2$, it will create additional maximum along the inflation trajectory. The solution reads~\footnote{There are other solutions; however, they introduce additional barriers between the inflation point and the minimum of the potential. Thus we will ignore them in this work.}:
\begin{equation}
\begin{split}
n=&2, \quad 0.1261< \beta <0.1268\,,\\
n=&3, \quad 0.2450< \beta <0.2454\,,\\
n=&4, \quad 0.3752< \beta <0.3755\,.\\
\end{split}
\end{equation}
Plugging the above solution into $C_6$, we find $C_6 \approx 350$ for $n=2$. As in this case $C_2,C_4 \ll C_6$, we find that this setup can generate inflation potential that agrees with CMB observations. We show a possible shape of the potential and its cross-section in figure~\ref{fig:model1-potential}. Theoretical predictions for this case can be found in figure~\ref{fig:inflationprediction} where $n=2$ is used as an example. It is straightforward to extend our results to $n>2$. We use different colors to label different choice of $C_4$. Solid and dashed lines represent results for $N_e=50$ and $N_e=60$, respectively. The latter cases are slightly shifted towards a smaller tensor-to-scalar ratio and lager spectral index. Increasing $C_4$ will decrease the spectral index and vice versa, as expected. As one can see from the figures, the spectral index $n_s$ can be in the $1\sigma$ region constrained by the CMB observation, while the tensor-to-scalar ratio $r$ is of order $\mathcal{O}(10^{-6})$. This is well below the current sensitivity. The running of spectral index $\alpha$~\footnote{See Appendix~\ref{app:SRconfig} for definition of the running of spectral index $\alpha$.} is of order $\mathcal{O}(-10^{-4})$, which might be testable in the CMB S4 mission~\cite{CMB-S4:2016ple, Munoz:2016owz} and future observations of 21 cm fluctuations~\cite{Kohri:2013mxa,Modak:2021zgb}.  We also show the parameter space spanned by $A(S,\Bar{S})$ and $\beta$ in figure~\ref{fig:parameterspace}. The left and right segments are results for $N_e=50$ and $N_e=60$, respectively. We use different color to represent different values of spectral index $n_s$.
%%%%%%%%%%%%%%%%%%%%%%%%%%%%%%%%%%%%%%%%%%%%%%%%%%%%%%%%%%

%%%%%%%%%%%%%%%%%%%%%%%%%%%%%%%%%%%%%%%%%%%%%%%%%%%%%%%%%%%%%%%%%%%%%%%
\subsubsection{$\mathcal{P}(j)=1 + \gamma\left(1-j/1728\right)^2$}
The next simple choice would be to turning on the $\phi^4$ contribution in $\mathcal{P}(j)$ by introducing a non-zero value of $\gamma$. In this case, $C_2$ remains unchanged while $C_4$ and $C_6$ get additional contribution, which we denote by $C_{4,\gamma}$ and $C_{6,\gamma}$:
\begin{equation}
\begin{split}
C_2 =& C_2|_{\gamma=0}\,,\\
C_4=&C_4|_{\gamma=0}+C_{4,\gamma}\\=&C_4|_{\gamma=0}-183.497\gamma+\frac{(437.766+9374.025n) \gamma}{A-3}\,,\\
C_6=&C_6|_{\gamma=0}+C_{6,\gamma}\\=&C_6|_{\gamma=0}+\gamma (1171.753n+1592.738)\\&+\frac{-269368.311\gamma^2-6821.952\gamma-113625.201n\gamma-74824.531n^2\gamma}{A-3}\,.
\end{split}
\end{equation}
Since $C_2$ is unchanged, the solution for $0<C_2<0.008$ remains the same as in the $\mathcal{P}(j)=1$ case:
\begin{equation}
3.596 + 12.771 n < A < 3.612 + 12.771 n\,.
\end{equation}
Again requiring $-2<C_4<5$, $\gamma$ is constrained to be:
\begin{equation}
\begin{split}
n=&2, \quad -0.119< \gamma <-0.106\,,\\
n=&3, \quad -0.287< \gamma< -0.274\,,\\
n=&4, \quad -0.529< \gamma< -0.516\,.\\
\end{split}
\end{equation}
These conditions ensure $C_2$ and $C_4$ are much smaller than $C_6$, so that the $\phi^2$ and $\phi^6$ terms dominate during inflation. We demonstrate the theoretical prediction for $n=2$ in figure~\ref{fig:inflationprediction}. There is no surprise that the prediction is similar to the previous case where we choose $\beta \neq 0$. Both cases have similar local expansion and inflation trajectory. The spectral index $n_s$ can again lie in the $1\sigma$ region constrained by the CMB observation, while the tensor-to-scalar ratio $r$ is of order $\mathcal{O}(10^{-6})$ and  the running of spectral index $\alpha$ is of order $\mathcal{O}(-10^{-4})$. We also show the parameter space spanned by $A(S,\Bar{S}), \gamma$ in figure~\ref{fig:parameterspace}. The left and right segments represent results for $N_e=50$ and $N_e=60$, respectively. We use different color to represent different values of spectral index $n_s$.

%%%%%%%%%%%%%%%%%%%%%%%%%%%%%%%%%%%%%%%%%%%%%%%%%%%%%%%%%%%%%%%%%%%%%%%
\subsubsection{$\mathcal{P}(j)=1 +\beta\left(1-j/1728\right) +\gamma\left(1-j/1728\right)^2$}
In this case, we can choose $A$ as a free parameter and deduce the value for $\beta$ and $\gamma$ accordingly. The expression for $C_{2n}$ now include the mixing terms between $\beta$ and $\gamma$, and they read:
\begin{equation}
\begin{split}
C_2 =&C_2|_{\beta=\gamma=0}+C_{2,\beta}\,,\\
C_4 =&C_4|_{\beta=\gamma=0}+C_{4,\beta}+C_{4,\gamma}- \beta \gamma \frac{28122.076}{A-3}\,,\\
C_6 =&C_6|_{\beta=\gamma=0}+C_{4,\beta}+C_{4,\gamma}+ \beta \gamma \left( -1757.630+\frac{314263.030n+603954.174}{A-3}\right)\,.
\end{split}
\end{equation}
For any arbitrary $A$ there are always two possible values of $\beta$ that ensure the quadratic coefficient $C_2$ sufficiently small:
\begin{equation}
\begin{split}
\beta_1 &\approx 0.016+0.333 n\,,\\
\beta_2 &\approx 0.094-0.026A+0.333n\,.
\end{split}
\end{equation}
The second solution encompasses the previous case with $ n=2,~\beta=0$ and $A=29.142$. One can then determine the appropriate $\gamma$ value to ensure $C_4 \ll C_6$. We find that the most favored region is $C_2\approx0.004$, $C_4<2$ and $C_6 \approx 900$. The theoretical prediction for $A=25$ and $n=2$ is shown in figure~\ref{fig:inflationprediction}. The physics of inflation remains nearly identical to the previous cases, yielding a similar prediction. The tensor-to-scalar ratio $r$ and the running of spectral index $\alpha$ are very small, while the spectral index $n_s$ remains within the $1\sigma$ region of the observational constraints. We use $A=25$ and $n=2$ as an example, some typical values of $\beta,\gamma$ and the corresponding predictions for inflation parameters are listed in table~\ref{app:inflationtable}.

One may wonder if it is possible to achieve successful inflation in the opposite direction, where the modular field slowly roll from $\tau=\omega$ to $\tau=i$. This scenario requires the potential to have a maximum at $\tau=\omega$ and a minimum at $\tau=i$. The latter condition can be easily fulfilled by choosing $m\neq 0$. However, satisfying the former condition necessitates $n=0$. In this case, $\tau=\omega$ is always a local minimum of the potential if $A>3$, regardless of the form of $\mathcal{P}(j)$.
This behavior can be seen from their local expansion:
\begin{equation}
\begin{split}
j(\phi) &\approx 1728\left[-9.36 (\phi-\phi_\omega)^3-1.56(\theta-\phi_\omega)^5\right]+\mathcal{O}\left((\phi-\phi_\omega)^6\right) \,,\\
V(\phi)&\approx V_0\left[1+ \frac{1}{2}\frac{A-2}{A-3}(\phi-\phi_\omega)^2+C_3(\phi-\phi_\omega)^3\right]+\mathcal{O}\left((\phi-\phi_\omega)^4\right)\,,
\label{eq:fullexpansionwithm}
\end{split}
\end{equation}
where $V_0=\frac{1728^{m}\Lambda^4\left[A(S,\bar{S})-3)\right]}{8\sin^3(2\pi/3)\abs{\eta(\omega)}^{12}}\abs{\mathcal{P}(0)}^2$ and $\phi_\omega=\sqrt{3/8}\,\ln3$ is the value of canonical field when $\tau=\omega$. Thus, as long as $A(S,\bar{S})>3$ (a condition necessary to ensure the potential remains positive during inflation), $\tau=\omega$ is always a local minimum of the potential. Consequently, we will not consider slow-roll inflation in this case.

%%%%%%%%%%%%%%%%%%%%%%%%%%%%%%%%%%%%%%%%%%%%%%%%%%%%%%%%%%%
\subsection{Slow roll along the left (or right) boundary}
We would extend our discussion to the case where $m,n\geq 2$. In this case, $\tau=i,\omega$ are both minima of the potential and we consider an inflation trajectory at the left boundary (or the imaginary axis) of the fundamental domain. Unlike the previous case where we have good understanding of both inflation point and minimum of the potential, we can not make any assumption where inflation will happen. Thus it is generally difficult to give analytic expressions so we will only show an example here. However, the treatment here is using the rich vacuum structure of the modular potential and it should be very generic.

Let's first find the canonical field at the left boundary. The kinetic term for imaginary part of $\tau$ reads:
\begin{equation}
{\cal L}_{\text{kin}}=\frac{1}{2}\frac{3}{2(\text{Im}\tau)^2}\left(\partial_\mu\text{Im}\tau\right)^2\,.
\end{equation}
One can make a field redefinition, $\text{Im}\tau=\exp(\sqrt{2/3}\, \phi)$, to introduce the canonical field $\phi$. The minimum of $\text{Im}\tau$ is $\sqrt{3}/2$, which corresponds to $\phi=\sqrt{3/2}\ln{(\sqrt{3}/2)}\approx-0.17$.

It has been noticed that there exist multiple local minima along the left boundary of the fundamental domain~\cite{Leedom:2022zdm}. In the specific case where $m=2,~n=2,~A(S,\Bar{S})=0,~\mathcal{P}(j)=1+10^{-3}j$, they find an additional AdS minimum at the left boundary. This minimum can be uplifted into a dS minimum by turning on the $A(S,\Bar{S})$ term. This was called Accidental Inflation in String Theory~\cite{Linde:2007jn}, where the up-lifting of adjacent minimum leads to inflation.
\begin{figure}[th]
\centering
\includegraphics[width=.50\linewidth]{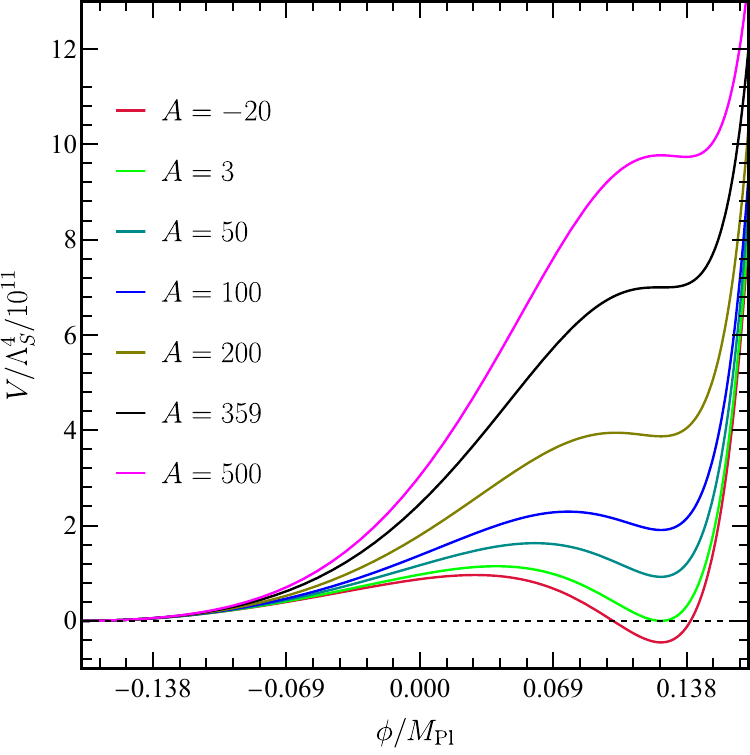}
\caption{When $m=2$, $n=2$ and $\beta=-0.633431$, we show the scalar potential at the left boundary of the fundamental domain across different values of the parameter $A$. Notably, we observe the emergence of almost flat potential when $A \sim \mathcal{O}(100)$. The flat plateau occurs around $\tau_I \approx 1.11$ or $\phi\approx0.13$. In addition,  the corresponding potential attains its minimum at $\tau=\omega$ when $A>3$. Particularly noteworthy is the case when $A=3$, where the vacuum becomes degenerate. Conversely, for $A<3$, such as $A=-20$, the AdS global minimum is located at $\tau=1.10714i$, while the fixed point $\tau=\omega$ serves as a Minkowski local minimum.}
\label{fig:Ascalarpotential}
\end{figure}

We exploit this idea and show how the potential are shaped by different $A(S,\bar{S})$ values. One can see from the figure~\ref{fig:Ascalarpotential}, when $A$ is small, this additional minimum located at $\phi\approx0.13$ is a (A)dS minimum, which is separated from the minimum at $\phi=\phi_\omega$ by a barrier. If one increases the value of $A$, the potential becomes flat and an inflection point will emerge. This very flat region in the potential may break up the normal slow roll inflation approximation and lead to so called ultra slow roll (USR) inflation~\cite{Kinney:2005vj,Martin:2012pe,Dimopoulos:2017ged}. It might enhance the curvature perturbation and leads to production of primordial black holes~\cite{Namjoo:2012aa,Mooij:2015yka, Germani:2017bcs}.

We find a narrow region where a SR-USR-SR transition occurs:
\begin{equation}
357.85<A<358.75\,.
\end{equation}
This case needs a very careful treatment and we leave it for further works.

\section{Summary and conclusion\label{sec:SaC}}

In this paper, we try to combine the ideas of modular stabilization and modular inflation. Inspired by the vacuum structure of the modular potential, we successfully find two different trajectories that can accommodate inflation phenomenology and agree with CMB observations. Those trajectories follow the boundaries of the fundamental domain and the property of modular symmetry plays a significant role in our construction. Modular symmetry ensures the flatness of the potential in the inflationary domain and the stabilization of the perpendicular field during inflation. We also find that modular symmetry is a very strong constraint that prevents us from approximating it as a simple hilltop inflation.

We have three different sets of parameters in our model. The first one is $A(S,\bar{S})$ in Eq.~\eqref{AVZ}, which determines the relative contribution from the dilaton sector. It has to be non-zero for successful inflation to happen. It is also the reason why we could evade some no-go theorem stated in previous studies on logarithmic K\"ahler potential inflation~\cite{Badziak:2008yg,Ben-Dayan:2008fhy,Covi:2008cn}. The second set of parameters is a pair of integers $(m,n)$ from the $H$ function in Eq.~\eqref{eq:H1}, which determines the vacuum structure of the inflation potentials. We have constructed different trajectories based on that. We keep $(m,n)$ as free parameters. They are relevant to the reheating process after inflation as they also affect how the inflaton oscillates at the minimum. This is not discussed in our paper, we leave it for further exploration. The third set of parameters $(\beta,\gamma)$ is coefficient of the $j$-invariant polynomial in Eq.~\eqref{eq:jpoly}. These coefficients actually shape the potential and are essential for constructing inflation potentials. 

In summary, the special properties of fixed points under modular symmetry motivate us to consider the following inflation scenario: when $n\geq2,~m=0$, we find the scalar potential near the fixed point $i$ can be flat enough to accommodate inflation. In this scenario, inflation occurs along the unit arc of the fundamental domain. It is necessary to include one of the parameters in $(\beta,\gamma)$, otherwise, the predicted spectral index $n_s$ is smaller than observed. When $n=2, m=2$, we find the possibility of realizing (ultra) slow-roll inflation by uplifting the adjacent minima of the potential. In this scenario, inflation occurs along the left boundary of the fundamental domain. The contribution from the dilaton sector $A(S,\bar{S})$ is important for such an uplifting. We do not consider the case when $m\geq 2,~n=0$, as both fixed points are local minima.

The fine-tuning problem still exists in our model. Once we fix the parameter $(m,n)$, $A(S,\bar{S})$ and $(\beta,\gamma)$ have to be fixed accordingly. There is no theoretical evidence, besides the argument for accidental inflation, suggesting that they should take these specific values. In this sense, we can only answer the question how a modular invariant inflation would look like. We can not ensure that modular stabilization naturally leads to inflation. In particular, we have followed a bottom-up approach to inflation instead of a top-down approach. Here we have assumed that the dilaton is stabilized. The stabilization of dilaton field during inflation could be achieved if the stabilization happens at a much higher energy scale than the inflation scale. The stabilization of the dilaton sector and its dynamics deserve further research. It is also interesting to consider a scenario in which the dilaton field serves as the inflaton field. 

In this paper, we have focused on the single-field inflationary scenario, selecting one degree of the modulus field to be the inflaton after normalization. We leave the consideration of the multi-field inflation scenario for possible future works. Moreover, string compactifications will lead to the appearance of abundant moduli fields. It is worth the effort to investigate how to construct the multiple moduli inflation model.

\acknowledgments

WBZ would like to thank Professor Manuel Drees and Yong Xu for fruitful discussions. WBZ would like to thank Professor Manuel Drees for carefully reading our draft. SYJ and GJD are supported by the National Natural Science Foundation of China under Grant Nos. 12375104, 11975224.

\appendix
\newpage

%%%%%%%%%%%%%%%%%%%%%%%%%%%%%%%%%%%%%%%%%%%%%%%%%%%%%%%%%%%%%%%%%%%%%%%%
\section{\label{app:modularforms}Modular symmetry and relevant modular forms}

The modular symmetry group denoted as $\Gamma$ is the group composed of two-dimensional matrices with integer entries and unit determinant~\cite{Feruglio:2017spp,Ding:2023htn}:
\begin{equation}
\Gamma=\Biggl\{
\begin{pmatrix}
a ~&~ b \\
c ~&~ d
\end{pmatrix}\Big|~ a,b,c,d \in \mathbb{Z},\quad ad-bc=1\Biggl\} \,.
\end{equation}
The complex modulus $\tau$ takes value in the upper half complex plane denoted as ${\cal H}=\{\tau \in \mathbb{C}~|~ {\rm Im}\tau >0\}$. The modular group $\Gamma$ acts on the complex modulus $\tau\in\mathcal{H}$ as the following linear fraction transformation,
\begin{equation}\label{eq:modularTrans}
\tau\rightarrow \gamma\tau=\frac{a\tau+b}{c\tau+d},~~~\gamma=\begin{pmatrix}
a  ~& b\\
c  ~& d
\end{pmatrix}\in\Gamma,~~\text{Im}\tau>0\,.
\end{equation}
One sees that $\gamma$ and $-\gamma$ give the same linear fraction transformation, and consequently the group of the above modular transformation is isomorphic to the projective group $\overline{\Gamma}\equiv\Gamma/\{\pm \mathds{1}\}$. If all elements of $\Gamma$ acting on any given point in ${\cal H}$, the point sets generated are equivalent. As a result, ${\cal H}$ can be decomposed as set of trajectories ${\cal H} /\Gamma$. The representative points of the trajectories $\mathcal{H} /\Gamma$ constitute the so-called fundamental domain of $\Gamma$ given by
\begin{equation}
{\cal D}=\{\tau \in {\cal H}~ \Big|~ |\tau|> 1, -1/2\leq {\rm Re(\tau)} < 1/2\}\cup\{|\tau|= 1,-1/2 \leq {\rm Re(\tau)} \leq  0\} \,,
\end{equation}
where no two points in ${\cal D}$ are related by modular transformations. We plot the fundamental domain in figure~\ref{fig:traj1}. The modular group $\Gamma$ has an infinite number of elements and it can be generated by two generators ${\cal S}$ and ${\cal T}$:
\begin{equation}
{\cal S}:\tau\rightarrow-\frac{1}{\tau}\,,~~~~~~ {\cal T}:\tau\rightarrow\tau+1\,.
\end{equation}
The matrix representation of $\mathcal{S}$ and $\mathcal{T}$ are given by
\begin{equation}
{\cal S}=
\begin{pmatrix}
~0 ~&~ 1\\
-1 ~&~ 0
\end{pmatrix}\,,\quad
{\cal T}=
\begin{pmatrix}
~1 ~&~ 1~\\
~0 ~&~ 1~
\end{pmatrix}\,,
\end{equation}
which obey the relations ${\cal S}^4=({\cal S}{\cal T})^3=1$ and $\mathcal{S}^2\mathcal{T}=\mathcal{T}\mathcal{S}^2$. In the fundamental domain $\mathcal{D}$, there are only three fixed points $\tau=i$, $\tau=e^{2\pi i/3}\equiv\omega$ and $\tau=i\infty$ which are invariant under $\mathcal{S}$, $\mathcal{ST}$ and $\mathcal{T}$ respectively. There is a class of complex functions over the plane ${\cal H}$ called modular forms, which are holomorphic functions of $\tau$ transforming under the modular groups as
\begin{equation}
f(\gamma\tau)=(c\tau+d)^k f(\tau),~~~\gamma\in\Gamma \,,
\end{equation}
where the weight $k$ is a generic non-negative integer. In the following, we report some modular functions involved in the analysis of scalar potential.

\label{app:modularforms}
\begin{itemize}
\item Dedekind eta function\\
The Dedekind eta function is a modular function of ``weight $1/2$''  defined as
\begin{equation}
\eta(\tau)=q^{1/24}\prod^{\infty}_{n=1}(1-q^n),\quad~~
q\equiv e^{2\pi i\tau}\,,
\end{equation}
which satisfies the identities $\eta(\tau+1)=e^{i\pi/12}\eta(\tau)$ and $\eta(-1/\tau)=\sqrt{-i\tau}\eta(\tau)$.
The $q$-expansion of eta function is given by
\begin{equation}
\eta=q^{1/24}[1-q-q^2+q^5+q^7-q^{12}-q^{15}+{\cal O}(q^{22})]\,.
\end{equation}
At the modular symmetry fixed points $\tau=\omega, i$, the eta function takes the following values,
\begin{equation}\label{eq:etaValue}
\eta(i) = \frac{\Gamma(1/4)}{2\pi^{3/4}}\,,~~~
\eta(\omega) = e^{-\frac{i\pi}{24}}\frac{3^{1/8}\;\Gamma^{3/2}(1/3)}{2\pi}\,.
\end{equation}

\item Eisenstein series\\
The Eisenstein series $G_{2k}(\tau)$ of weight $2k$ for integer $k\geq 1$ are defined as~\cite{DHoker:2022dxx}
\begin{equation}\label{eq:G2kE2k}
G_{2k}(\tau)=2\zeta(2k)+2\frac{(2\pi i)^{2k}}{(2k-1)!}\sum_{n=1}^{\infty}\sigma_{2k-1}(n)q^n\,,
\end{equation}
where $\zeta(z)$ denotes the Riemann's zeta function, and $\sigma_p(n)$ is the sum of the $p$th power of the divisors of $n$,
\begin{equation}
\sigma_{p}(n)=\sum_{d|n}d^p\,,
\end{equation}
where $d|n$ is shorthand for ``$d$ divides $n$''. The Eisenstein series $G_{2k}(\tau)$ are modular forms of weight $2k$ for any integer $k>1$. However, $G_2(\tau)$ is not a modular form and its transformation is given by
\begin{equation}
G_2(\gamma\tau)=(c\tau+d)^2G_2(\tau)-2\pi i c(c\tau + d)\,.
\end{equation}
Note the modular functions $G_2(\tau)$ and the Dedekind eta are related by
\begin{equation}
\frac{\partial_\tau\eta(\tau)}{\eta(\tau)}=\frac{i}{4\pi}G_2(\tau)\,.
\end{equation}
The modified weight $2$ Eisenstein series $\widehat{G}_2$ is defined by
\begin{equation}\label{eq:G2hat}
\widehat{G}_2(\tau)=G_2(\tau) + \frac{2\pi}{i(\tau-\bar{\tau})}\,,
\end{equation}
which is a non-holomorphic function but preserves modularity. Consequently $\widehat{G}_2(\tau)$ vanishes at the fixed points:
\begin{equation}\label{eq:widehatG2atfix}
\widehat{G}_2(i)=\widehat{G}_2(\omega)=0\,.
\end{equation}
Sometimes it is more convenient to work with the normalized  Eisenstein series $E_{2k}(\tau)$, which differ from $G_{2k}$ in the normalization constant~\cite{DHoker:2022dxx}:
\begin{equation}\label{eq:GtoE}
E_{2k}(\tau)=\frac{G_{2k}(\tau)}{2\zeta(2k)}\,.
\end{equation}
We provide some values of $E_{2k}(\tau)$ relevant to this paper:
\begin{eqnarray}
E_2(i) &=& \frac{3}{\pi}\,,\quad\quad\quad\quad~ E_2(\omega) = \frac{2\sqrt{3}}{\pi}\,,\\
E_4(i) &=& \frac{3\Gamma^8(1/4)}{(2\pi)^6} \,,~~~~
E_4(\omega) = 0\,, \\
E_6(i) &=& 0\,,\quad\quad\quad\quad\quad
E_6(\omega) = \frac{6^3\Gamma^{18}(1/3)}{(2\pi)^{12}}\,.
\end{eqnarray}
As regards the derivatives of the Eisenstein series, Ramanujan-Shen's differential equation~\cite{kobayashi2024ramanujan} is useful:
\begin{equation}
qE'_{2k-2}=\frac{k-1}{2\pi^2\zeta(2k-2)}\sum_{n=1}^{k-1}\zeta(2n)\zeta(2k-2n)(E_{2n}E_{2k-2n}-E_{2k})\,,~~\text{for}~~k\geq2\,,
\end{equation}
where $E'_{2k-2}=\text{d}E_{2k-2}/\text{d}q$. The first several relations read:
\begin{eqnarray}
\nonumber qE'_2(q) &=&\frac{1}{12}(E^2_2-E_4)\,,\\
\nonumber qE'_4(q) &=&\frac{1}{3}(E_2E_4-E_6)\,, \\
\label{eq:Ramanujan} qE'_6(q) &=&  \frac{1}{2}(E_2E_6-E^2_4)\,.
\end{eqnarray}
Using the first identity in Eq.~\eqref{eq:Ramanujan} and
$\text{d}/\text{d}\tau=2\pi i q\,\text{d}/\text{d} q$, we find the derivative of $\widehat{G}_2$ is as follows:
\begin{equation}
\partial_\tau \widehat{G}_2 (\tau) = \frac{i\pi^3}{18}(E^2_2-E_4)-\frac{2\pi}{i(\tau -\bar{\tau})^2}\,,
\end{equation}
and the numerical values of $\partial_\tau \widehat{G}_2(\tau)$ at the fixed points are
\begin{equation}
\partial_\tau \widehat{G}_2(i) = -i\frac{\Gamma^8(1/4)}{384\pi^3}\,,\quad\partial_\tau \widehat{G}_2 (\omega) = 0 \,.
\end{equation}
The anti-holomorphic derivative of $\widehat{G}_2(\tau,\bar{\tau})$ is
\begin{equation}
\quad\partial_{\bar{\tau}} \widehat{G}_2  (\tau) = -\partial_{\tau} \overline{\widehat{G}}_2 (\tau) = \frac{2\pi}{i(\tau-\bar{\tau})^2}\,,
\end{equation}
the corresponding numerical values at the fixed points are
\begin{equation}\label{eq:pG2hat}
\begin{split}
\quad\partial_{\bar{\tau}} \widehat{G}_2  (\tau)\Big|_{\tau=i} &= -\partial_{\tau} \overline{\widehat{G}}_2 (\tau) \Big|_{\tau=i}=i\frac{\pi}{2}\,, \\
\quad\partial_{\bar{\tau}} \widehat{G}_2  (\tau)\Big|_{\tau=\omega} &= -\partial_{\tau} \overline{\widehat{G}}_2 (\tau) \Big|_{\tau=\omega}=i\frac{2\pi}{3}\,.
\end{split}
\end{equation}
These relations are useful when calculating the derivatives of the scalar potential.

\item{Klein $j$-invariant function}

The Klein $j$-invariant function is a modular form of weight zero, defined in terms of Dedekind eta function and Eisenstein series as follows~\cite{Novichkov:2022wvg,Cvetic:1991qm}
\begin{equation}\label{eq:jG4}
j(\tau)\equiv\frac{3^6 5^3}{\pi^{12}}\frac{G^3_4(\tau)}{\eta^{24}(\tau)}=\frac{3^6 5^3}{\pi^{12}}\frac{G^3_4(\tau)}{\Delta(\tau)}\,,\quad \Delta(\tau)\equiv\eta^{24}(\tau)\,,
\end{equation}
which implies
\begin{equation}
\label{eq:j-1728}
j(\tau)-1728 =\left(\frac{945}{2\pi^6}\right)^2\left(\frac{G_6(\tau)}{\eta^{12}(\tau)}\right)^{2} = \left(\frac{945}{2\pi^6}\right)^2\frac{G_6^2(\tau)}{\Delta(\tau)}\,.
\end{equation}
From Eqs.~(\ref{eq:jG4}, \ref{eq:j-1728}), we can see that the two expressions of $H$ function in Eq.~\eqref{eq:H1} and Eq.~\eqref{eq:H2} are equivalent. Given the identity of Eq.~\eqref{eq:GtoE}, $j$ and $j-1728$ can also be compactly written as
\begin{equation}
\label{eq:j-E4E6}j(\tau)-1728 = \left(\frac{E_6(\tau)}{\eta^{12}(\tau)}\right)^2\,,\quad
j(\tau) = \left(\frac{E_4(\tau)}{\eta^8(\tau)}\right)^3\,.
\end{equation}

This $j$-function is a one-to-one map between points in the fundamental domain and the whole complex plane. At the fixed points, one has
\begin{equation}\label{eq:jnumvalue}
j(i\infty)=+\infty\,,\quad\quad j(\omega)=0\,,\quad\quad j(i)=1728=12^3\,.
\end{equation}
For convenience, the $q$-expansion of $j$-function is given by
\begin{eqnarray}
\nonumber j(\tau)  &=&  744 + \frac{1}{q} + 196884 q + 21493760 q^2 + 864299970 q^3 +
 20245856256 q^4\\
 && + 333202640600 q^5
 + 4252023300096 q^6 +
 44656994071935 q^7 + {\cal O}(q^8)~~\,.
\end{eqnarray}
The derivatives of the $j$-function read as,
\begin{eqnarray}
\nonumber\frac{\partial j}{\partial \tau} &=& -2\pi i \frac{ E_6 (\tau) E^2_4(\tau)}{\eta^{24}(\tau)}\,,\\
\label{eq:pj1}\frac{\partial^2 j}{\partial \tau^2}&=& \frac{(2\pi)^2}{\eta^{24}(\tau)}\left[\frac{1}{6}E_2(\tau)E_4^2(\tau) E_6(\tau)-\frac{1}{2}E^4_4(\tau) - \frac{2}{3}E^2_6(\tau)E_4(\tau)\right]\,,
\end{eqnarray}
which are useful to calculate the second derivative of scalar potential. The derivative of $j$-function also vanishes at the fixed points:
\begin{equation}
\frac{\partial j}{\partial \tau}\Big|_{\tau=i} =  \frac{\partial j}{\partial \tau}\Big|_{\tau=\omega} = 0\,.
\end{equation}

\item{$H$ function}\\
Using Eq.~\eqref{eq:j-E4E6}, the $H$ function in Eq.~\eqref{eq:H1} can also be written as,
\begin{equation}
H(\tau)=\left(\frac{E_6(\tau)}{\eta^{12}(\tau)}\right)^{m}\left(\frac{E_4(\tau)}{\eta^{8}(\tau)}\right)^{n}\mathcal{P}(j(\tau))\,.
\end{equation}
The numerical values of the $H$ function at the fixed points are determined to be
\begin{equation}\label{eq:hfunc-mn}
\begin{split}
H(i) &=
\begin{cases}
0&  \quad m>0\\
12^n{\cal P}(1728) &  \quad   m=0
\end{cases}\,,\\
H(\omega) &=
\begin{cases}
0&  \quad n>0\\
i^m 2^{3m} 3^{\frac{3m}{2}}{\cal P}(0) &  \quad   n=0
\end{cases}\,.
\end{split}
\end{equation}
The derivative of the $H$ function is found to be:
\begin{equation}\label{eq:pH1}
\partial_\tau H = -i\pi H \left( m \frac{E^2_4}{E_6} + \frac{2n}{3} \frac{E_6}{E_4}  + \frac{i}{\pi}\frac{\text{d} \ln {\cal P}}{\text{d} j}\frac{\text{d}j}{\text{d} \tau}\right)\,,
\end{equation}
and the corresponding numerical values at the fixed points are:
\begin{equation}\label{eq:D1hfunc-mn}
\begin{split}
\partial_\tau H (\tau)\Big|_{\tau=i}=&
\begin{cases}
0& \quad m=0~\textrm{or} ~m>1\\
-i 2^{2n+2}3^{2+n}\frac{\Gamma^4(1/4)}{(2\pi)^2}{\cal P}(1728) &  \quad   m=1
\end{cases}\,,\\
\partial_\tau H (\tau)\Big|_{\tau=\omega}=&
\begin{cases}
0&  \quad n=0~\textrm{or} ~n>1\\
-i^{m+1}  2^{3m+3}3^{3m/2+1}e^{i\frac{\pi}{3}}(2\pi)^{-3}\Gamma^6(1/3){\cal P}(0) & \quad   n=1
\end{cases}\,.
\end{split}
\end{equation}
When determining the Hessian matrix of the scalar potential, the second order derivative of the $H$ function is needed. It is found that $\partial^2_\tau H$ takes the following value at the fixed points:
\begin{equation}\label{eq:D2hfunc-mn}
\begin{split}
\partial^2_\tau H (\tau)\Big|_{\tau=i} =&
\begin{cases}
0&  \quad m>2\\
- 3^{n+4} 2^{5+2n} \frac{\Gamma^8(1/4)}{(2\pi)^4}{\cal P}(1728) &  \quad   m=2 \\
3^{n+2}2^{2n+2}\frac{\Gamma^4(1/4)}{(2\pi)^2}{\cal P}(1728)&  \quad   m=1 \\
-2^{2n-1}3^n\frac{\Gamma^8(1/4)}{(2\pi)^4}{\cal P}(1728)\left(n+3^4 2^6 \frac{  {\cal P}'(1728)}{ {\cal P}(1728) } \right)& \quad   m=0\\
\end{cases}\,,\\
\partial^2_\tau H (\tau)\Big|_{\tau=\omega} =&
\begin{cases}
0&  \quad n=0~\textrm{or}~n>2 \\
-i^{m}e^{i\frac{2\pi}{3}} 2^{3m+7} 3^{\frac{3m}{2}+2} \frac{\Gamma^{12}(1/3)}{(2\pi)^6}{\cal P}(0) &  \quad   n=2\\
i^{m}e^{i\frac{\pi}{3}} 2^{3m+4} 3^{\frac{3}{2}m + \frac{1}{2}} \frac{\Gamma^6(1/3)}{(2\pi)^3}{\cal P}(0)  & \quad n=1
\end{cases}\,.
\end{split}
\end{equation}
\end{itemize}

%%%%%%%%%%%%%%%%%%%%%%%%%%%%%%%%%%%%%%%%%%%%%%%%%%%%%%%%%%%%%%%%%%%%%%%%
\section{Parameters of slow roll inflation and current bounds \label{app:SRconfig}}

Inflation has been widely employed to address many shortcomings of the standard big bang cosmological model, such as the horizon and flatness problems. It suggests a period of exponential expansion, during which quantum fluctuations seed the primordial perturbations in our universe. The simplest way to realize inflation is through the slow roll mechanism: The inflationary field, inflaton $\phi$, evolves over a flat region of the potential $V(\phi)$~\cite{Liddle:1994dx}. The potential energy sources the exponential expansion, and the flatness of the potential ensures that this expansion lasts for a sufficiently long duration. Conventionally, we use  derivatives of the potential to measure its flatness, which are referred to as the first (second, third, fourth) slow roll parameters~\cite{Forconi:2021que}:
\begin{eqnarray}
\label{vare}  \varepsilon_V &=& \frac{M_{\text{Pl}}^2}{2}\left(\frac{V'}{V}\right)^2 \,,\\
\label{etaV}  \eta_V &=& M_{\text{Pl}}^2\left(\frac{V''}{V}\right)\,,\\
\label{xiV}   \xi_V &=& M_{\text{Pl}}^4\left(\frac{V'V'''}{V^2}\right)^{\frac{1}{2}}\,,\\
\label{varpiV}    \varpi_V  &=& M_{\text{Pl}}^6\left(\frac{V'^2V^{''''}}{V^3}\right)^{\frac{1}{3}}\,,
\end{eqnarray}
where $V$ represents the potential , and the prime $'$ denotes the derivative of the potential with respect to the inflaton field $\phi$. The duration of the expansion is reflected by the growth of the scale factor $a$. We define the number of e-folds for a given mode $k_*$ as the logarithm of its growth:
\begin{equation}\label{eq:Nint}
N_e(\phi_*)= \ln{\left(\frac{a_{\textrm{end}}}{a_*}\right)} =\int_{\phi_*}^{\phi_{\text{end}}}\frac{\text{d}\phi}{\sqrt{2\varepsilon_V}}\,,
\end{equation}
where $a_*$ represents the scale factor when the $k_*$ mode first crosses out of the horizon, and $a_{\textrm{end}}$ is the scale factor at the end of inflation. Here $\phi_{*}$ and $\phi_{\text{end}}$ are their corresponding field values respectively. Successful inflation requires both the first and second slow-roll parameters to be small, $\varepsilon_V,\abs{\eta_V} \ll 1$, and inflation terminates when they become of order one: $\varepsilon_V(\phi_{\text{end}})=1$ or $|\eta_V(\phi_{\text{end}})|=1$.

The constraints on inflation phenomenology mainly arise from the cosmic microwave background (CMB). The overall isotropies of the CMB tell us that the number of e-folds must be sufficiently large. Its small anisotropies characterize the primordial cosmological perturbations. For single-field inflation, the reduced spectrum of curvature perturbations is usually parameterized by a power law:
\begin{equation}
\mathcal{P_R}(k)=A_s\left(\frac{k}{k_*}\right)^{n_s-1}\,,
\end{equation}
where $k_*$ serves as a reference (or ``pivot'') scale. The spectral index, denoted by $n_s$, is given by $n_s - 1  = 2\eta_V(\phi_*)-6\varepsilon_V(\phi_*)$ for single-field slow-roll inflation. The running of spectral index $\alpha=\textrm{d}n_s/\textrm{d}\log{k}$, which describes the scaled dependence of spectral index, can also be calculated from derivatives of the potential: $\alpha = 16 \epsilon_V \eta_V -24 \epsilon_V^2-2\xi_V^2$.

The tensor perturbations induce primordial gravitational waves, exhibiting a similar power-law behavior:
\begin{equation}
\mathcal{P}_t(k)=A_t\left(\frac{k}{k_*}\right)^{n_t}.
\end{equation}
The tensor-to-scalar ratio $r$ is defined as $r=A_t/A_s$. In single-field inflation, the tensor spectral index $n_t=-r/8=-2\varepsilon_V(\phi_*)$, which is know as the consistency relation. The current bounds on the spectral index and tensor-to-scalar ratio are important for constraining inflationary models:~\cite{Planck:2018jri,BICEP:2021xfz}:
\begin{equation}
\begin{split}
\ln{(10^{10}A_s)} &= 3.044 \pm 0.014\, (68 \% \textrm{CL})\,,\\
n_s &= 0.9649 \pm  0.0042 \, (68 \% \textrm{CL})\,,\\
\alpha &= 0.0045\pm0.0067\, (68 \% \textrm{CL})\,,\\
r & <0.036\, (95\% \textrm{CL}) \,.
\end{split}
\label{eq:PlanckResults}
\end{equation}
The detailed bound on the number of e-folds depends on the post-inflationary dynamics~\cite{Liddle:2003as,Martin:2010kz}, and it is mostly chosen to be $50<N_e<60.$

%%%%%%%%%%%%%%%%%%%%%%%%%%%%%%%%%%%%%%%%%%%%%%%%%%%%%%%%%%%%%%%%
\section{Concrete examples of modular inflation\label{app:MR}}
In this section we list some concrete examples where modular inflation is realized with different choices of parameters and their predictions, as shown in table~\ref{tab:n2m0casechose2}. All the cases share same features: The tensor to scalar ratio $r$ is smaller than $10^{-6}$, while the running of spectral index $\alpha \approx -2\xi_V^2$ is of order $\mathcal{O}(-10^{-4})$. It is hard to detect such a small tensor to scalar ratio $r$ but we might have enough sensitivity on $\alpha$ in the future CMB S4 mission and observation of 21 cm fluctuations~\cite{CMB-S4:2016ple, Munoz:2016owz,Kohri:2013mxa,Modak:2021zgb}.
\begin{table}[hptb!]
\centering
\resizebox{0.8\textwidth}{!}{
\begin{tabular}{c||c c c c}\hline
$N_e$ &\multicolumn{4}{c}{50}  \\\hline\hline
$A$   &$29.1470$  ~&~ $24.3060$ ~&~ $29.1495$ ~&~ $25$  \\
$\beta$ & $0$  ~&~ $0.126434$ ~&~ $0$ ~&~ ~~$0.108376$   \\
$\gamma$ &$0$   ~&~ $0$ ~&~ $-0.115590$ ~&~ $-0.016932$   \\\hline
$n_s$  &\textcolor{red}{$0.9462$} & $0.9637$ ~&~ $0.9643$ ~&~ $0.9639$   \\
$\log_{10}r$ &$-7.39$ ~&~ $-6.03$ ~&~ $-6.30$ ~&~ $-6.04$   \\
$\varepsilon_V$  &$2.5\times 10^{-9}$  & $6.7\times 10^{-8}$ ~&~ $3.1\times 10^{-8}$  ~&~ $5.6\times 10^{-8}$  \\
$\eta_V$ & $-0.0269$  ~&~  $-0.019$ ~&~ $-0.018$ ~&~  $-0.018$   \\
$\xi^2_V$  &$0.00054$ ~&~ $0.00028$ ~&~ $0.00035$ ~&~ $0.00033$   \\
$\varpi^2_V$ &$-7.7\times 10^{-6}$  &  $-1.2\times 10^{-5}$ ~&~ $-1.2\times 10^{-6}$ ~&~ $-1.2\times 10^{-5}$   \\\hline\hline
$N_e$ &\multicolumn{4}{c}{55}     \\\hline\hline
$A$   & $29.1470$ ~&~ $24.3091$ ~&~ $29.1441$ ~&~ $25$  \\
$\beta$ & $0$  & $0.126425$ ~&~ $0$ ~&~ ~~$0.108152$  \\
$\gamma$ & $0$ & $0$ ~&~ $-0.115640$ ~&~ $-0.017168$   \\\hline
$n_s$  & \textcolor{red}{$0.9510$} ~&~ $0.9649$ ~&~ $0.9649$ ~&~   $0.9650$ \\
$\log_{10}r$ & $-7.50$ & $-6.04$ ~&~ $-6.60$ ~&~ $-6.42$  \\
$\varepsilon_V$  & $1.9\times 10^{-9}$  ~&~ $5.7\times 10^{-8}$ ~&~  $1.6\times 10^{-8}$ ~&~ $1.5\times 10^{-8}$ \\
$\eta_V$ &$-0.024$  ~&~  $-0.018$ ~&~ $-0.018$ ~&~ $-0.018$   \\
$\xi^2_V$  & $0.00045$ ~&~ $0.00021$ ~&~ $0.00038$ ~&~ $0.00037$   \\
$\varpi^2_V$ & $-5.9\times 10^{-6}$ & $-8.6\times 10^{-6}$ ~&~ $-8.5\times 10^{-6}$ ~&~  $-7.8\times 10^{-6}$  \\\hline\hline
$N_e$ & \multicolumn{4}{c}{60}    \\\hline\hline
$A$   & 29.1470&24.3108 ~&~ 29.1548 ~&~ 25  \\
$\beta$ &$0$  & 0.126420 ~&~ $0$ ~&~ 0.108104   \\
$\gamma$ &$0$  & $0$ ~&~ $-0.115567$ ~&~ $-0.017219$   \\\hline
$n_s$  &\textcolor{red}{0.9551} & 0.9649  ~&~ 0.9649  ~&~ 0.9654  \\
$\log_{10}r$ &$-7.61$ & $-6.09$ ~&~ $-6.34$ ~&~ $-6.61$   \\
$\varepsilon_V$  & $1.5\times 10^{-9}$ & $5.1\times 10^{-8}$ ~&~ $2.8\times 10^{-8}$ ~&~ $1.54\times 10^{-8}$  \\
$\eta_V$ & $-0.0225$  &  $-0.018$ ~&~ $-0.018$ ~&~ $-0.017$   \\
$\xi^2_V$  & $0.00037$ & $0.00013$ ~&~ $0.00012$ ~&~ $ 0.00033$   \\
$\varpi^2_V$ & $-4.7\times 10^{-6} $  & $-5.9\times 10^{-6}$ ~&~ $-5.8\times10^{-6}$  ~&~ $-5.4\times 10^{-6}$   \\\hline
\end{tabular}}
\caption{\label{tab:n2m0casechose2} We present numerical results of the slow-roll parameters $\{\varepsilon_V,\eta_V,\xi^2_V,\varpi^3_V\}$ and inflationary predictions $n_s$ and $r$ for various combinations of $A$, $\beta$ and $\gamma$ in the inflation potential. Notably, we highlight the results for distinguished values of $A$. It's worth noting that the spectral index $n_s$ is a bit small in the cases of $\beta=\gamma=0$, as indicated by data plotted in red. }
\label{app:inflationtable}
\end{table}

%%%%%%%%%%%%%%%%%%%%%%%%%%%%%%%%%%%%%%%%%%%%%%%%%%%%%%%%%%%%%%%%%%%%%%

\section{A toy model of dilaton stabilization \label{app:dilatonstablization}}

In this paper we have assumed that dilaton field $S$ is stabilized while the modulus field $\tau$ remains dynamic. We would like to justify this assumption using a toy model, where stabilization of dilaton filed does not constraint the modulus potential. This toy model is inspired by the racetrack scenario, which has been used to study moduli stabilization and inflation~\cite{Krasnikov:1987jj,Banks:1995dp,Barreiro:1998aj,Blanco-Pillado:2004aap,Blanco-Pillado:2006dgl}. Here we consider a two-component superpotential with different energy scales: 
\begin{equation}
\mathcal{W} = \Lambda_1 \Delta(S)g(\tau)+\Lambda_2 \Omega(S)f(\tau) 
\end{equation}
where $g(\tau)$ and $f(\tau)$ are  weight $-3$ modular functions, both of them can be parameterized by $H(\tau)\eta^{-6}(\tau)$ with different $m,n$ and polynomial $\mathcal{P}(j(\tau))$. $\Delta(S)$ and $\Omega(S)$ are holomorphic functions of $S$. The second term $\Lambda_2\Omega(S)f(\tau)$ was used in our paper to generate the inflation potential. $\Lambda_2$ is the inflation scale and $\Lambda_1$ refers the mass scale of dilaton field, which can be much larger than the inflation scale $\Lambda_1 \gg \Lambda_2$. Then it would be the first term dominate the potential while the second term only perturb it. One can suppose the first term determines the minimum of $S$. If the dilaton field is stabilized at $S_0$ with the following property:
\begin{equation}
\Delta(S_0)=\Delta_S(S_0)=0,\quad \Omega_S(S_0)+K_S \Omega(S_0)\neq 0 \,\,,
\end{equation}
the full potential 
\begin{eqnarray}
\nonumber V &=& e^K \Lambda_1^2 \left[K^{S \bar{S}}|\Delta_S(S) g(\tau) + K_S \Delta(S)g(\tau)|^2+\abs{\Delta(S)}^2K^{\tau \bar{\tau}}D_\tau g(\tau) D_{\Bar{\tau}}\overline{g(\tau)}-3|\Delta(S)g(\tau)|^2\right]\\
\nonumber &+&e^K \Lambda_2^2\left[K^{S \bar{S}}|\Omega_S(S) f(\tau) + K_S \Omega(S)f(\tau)|^2+\abs{\Omega(S)}^2K^{\tau \bar{\tau}}D_\tau f(\tau) D_{\Bar{\tau}}\overline{f(\tau)}-3|\Omega(S)f(\tau)|^2\right]\\
&+&\textrm{cross terms}\,,
\end{eqnarray}
would reduce to the potential used in the paper at the minimum of $S$:
\begin{eqnarray}
\nonumber V(S_0,\tau) &=&  e^K \Lambda_2^2 \Big[K^{S \bar{S}}|\Omega_S(S) f(\tau) + K_S \Omega(S)f(\tau)|^2+\abs{\Omega(S)}^2K^{\tau \bar{\tau}}D_\tau f(\tau) D_{\Bar{\tau}}\overline{f(\tau)}\\
&&-3|\Omega(S)f(\tau)|^2\Big]\Big|_{S=S_0}\,.
\end{eqnarray}
Phenomenally, this means the mass of dilaton is much larger than the energy scale of inflation. The influence of inflaton to dilaton stabilization would be suppressed by $\Lambda_2/\Lambda_1$. 
\flushbottom
\newpage
%\bibliographystyle{JHEP}
%\bibliography{ref}
%\end{document}

\providecommand{\href}[2]{#2}\begingroup\raggedright\endgroup

\end{document}